\documentclass[aps,floatfix,superscriptaddress,longbibliography,prl,reprint]{revtex4-2}

\newcommand{\papertitle}{Moiré-enabled polariton control in large-angle twisted bilayer graphene}

\usepackage[utf8]{inputenc}
\usepackage{graphicx}
\usepackage{wrapfig}
\usepackage{amsmath}
\usepackage{amssymb}
\usepackage{caption}
\usepackage{natbib}
\usepackage{float}
\usepackage{tikz}
\usepackage{stfloats}
\usepackage{siunitx}
\usepackage{dcolumn}
\usepackage{lmodern}
\usepackage[T1]{fontenc}
\usepackage[unicode=true,
	    colorlinks=true,
	    linkcolor=blue,
	    citecolor=blue,
	    urlcolor=blue]{hyperref} 
\usepackage{sansmath}
\usepackage{blindtext}
\usepackage[normalem]{ulem}

\usepackage{chemformula}
\usepackage{empheq}
\usepackage{comment}
\usepackage{bm}
\usepackage{gensymb}

\usepackage{caption}
\usepackage[misc]{ifsym}
\usepackage{fontawesome}

\usepackage{lineno}

\setcitestyle{super}

\usepackage{xr}
\makeatletter

\newcommand*{\addFileDependency}[1]{
\typeout{(#1)}

\@addtofilelist{#1}
\IfFileExists{#1}{}{\typeout{No file #1.}}
}\makeatother

\newcommand*{\myexternaldocument}[1]{%
\externaldocument{#1}%
\addFileDependency{#1.tex}%
\addFileDependency{#1.aux}%
}

\myexternaldocument{supplement}

\usepackage[top=2.5cm,bottom=2.5cm,left=1.5cm,right=1.5cm]{geometry}

\begin{document}

\title{\Large\textsf{\papertitle}}

\author{Bogdan Borodin}
\email{bborodin@nd.edu}
\affiliation{\footnotesize Department of Physics and Astronomy, University of Notre Dame, Notre Dame, IN 46556,~USA}
\affiliation{\footnotesize Stavropoulos Center for Complex Quantum Matter, University of Notre Dame, Notre Dame, IN 46556,~USA}

\author{Xuanpu Zhang}
\affiliation{\footnotesize Department of Physics and Astronomy, University of Notre Dame, Notre Dame, IN 46556,~USA}
\affiliation{\footnotesize Stavropoulos Center for Complex Quantum Matter, University of Notre Dame, Notre Dame, IN 46556,~USA}

\author{Christopher Yang}
\affiliation{\footnotesize Department of Physics and Astronomy, University of California, Irvine, California 92697, ~USA}

\author{Kenji Watanabe}
\affiliation{\footnotesize Research Center for Electronic and Optical Materials, National Institute for Materials Science, 1-1 Namiki, Tsukuba 305-0044,~Japan}

\author{Takashi Taniguchi}
\affiliation{\footnotesize Research Center for Materials Nanoarchitectonics, National Institute for Materials Science,  1-1 Namiki, Tsukuba 305-0044,~Japan}

\author{Cyprian Lewandowski}
\email{clewandowski@magnet.fsu.edu}
\affiliation{Department of Physics, Florida State University, Tallahassee, Florida 32306, USA}
\affiliation{National High Magnetic Field Laboratory, Tallahassee, Florida 32310, USA}

\author{Petr Stepanov}
\email{pstepano@nd.edu}
\affiliation{\footnotesize Department of Physics and Astronomy, University of Notre Dame, Notre Dame, IN 46556,~USA}
\affiliation{\footnotesize Stavropoulos Center for Complex Quantum Matter, University of Notre Dame, Notre Dame, IN 46556,~USA}


\begin{abstract}

Twisted bilayer graphene (TBG) is commonly viewed through two limiting regimes: strongly correlated flat bands near the magic angle and nearly decoupled Dirac cones at large twist angles. Here we explore the intermediate regime. Using hyperbolic plasmon--phonon polaritons (HP$^{3}$) to probe the dynamical electronic response, we investigate bilayer graphene from AB stacking to $5.5^\circ$ TBG. We identify a narrow density window in which the HP$^{3}$ wavelength becomes strongly gate tunable. The window opens when doping suppresses Landau damping and a coherent graphene plasmon emerges. It closes at a second crossover, which appears only above $4^\circ$ and shifts systematically with twist angle. Continuum-model and electrodynamic calculations trace this second crossover to hybridization between the central and remote moiré bands, which weakens the density dependence of the Drude response. The resulting polaritonic tunability is concentrated within a twist-angle-selected density window and produces total wavelength changes approaching $90\%$. Our results demonstrate that remote moiré bands reshape the collective electrodynamics of TBG far from the magic-angle regime and establish large-angle TBG as a platform for gate-reconfigurable mid-infrared polaritons and material cavities.
\end{abstract}

\maketitle
The electronic behavior of twisted bilayer graphene (TBG) is often discussed in terms of two contrasting limits. Near the magic angle, narrow bands promote strong correlation effects\cite{cao2018unconventional, andrei2020graphene,bistritzer2011moire}, whereas at larger twist angles the two graphene layers behave increasingly like independent Dirac systems, with interactions playing a much less prominent role\cite{lopes2007graphene, luican2011single, trambly2016electronic, jones2020observation, dale2023layer}. While this picture captures many low-energy transport phenomena, theory predicts that remote bands remain strongly hybridized by the moiré potential even at large twist angles, giving rise to minibands, modified optical conductivity, and new collective excitations\cite{stauber2013optical, hamer2022moire, tabert2013optical, moon2013optical, cavicchi2024theory}. Large-angle TBG thus offers a route to moiré physics beyond the flat-band paradigm, in which band hybridization remains important without relying on the fragile many-body phases of the magic-angle regime\cite{andrei2020graphene, hamer2022moire}. 

The collective optical response of this regime is relevant not only as a probe of moiré electronic structure\cite{papaj2023probing, li2024infrared, cui2024nanoscale}, but also as a possible resource for engineering the electromagnetic environment of quantum materials\cite{kuang2024optical, stauber2016quasi, woessner2015highly, schlawin2022cavity}. Recent experiments have shown that confined hyperbolic modes of hBN can couple resonantly to material excitations and alter the superconducting response of an adjacent molecular crystal\cite{keren2026cavity}, motivating the development of material cavities whose spectra can be controlled without continuous optical driving\cite{yang2023optical}. In this context, moiré superlattices offer an additional design parameter: twist angle and electrostatic gating can reshape the electronic conductivity and thereby modify the wavelength, confinement, and losses of hybrid polariton modes\cite{ni2015plasmons, sunku2018photonic, lewandowski2019intrinsically, papaj2020plasmonic, hesp2021observation}. Establishing how remote moiré bands control these modes is therefore important both for spectroscopic access to large-angle TBG and for the longer-term goal of constructing actively reconfigurable material cavities\cite{huang2022observation, herzig2024high}.

Direct measurements of this collective response are nevertheless challenging\cite{hesp2021nano, barbosa2025ultra}. Transport primarily probes states near the Fermi level in the central bands of TBG and is often limited by the achievable carrier density. The central bands are the two moiré bands adjacent to charge neutrality, which become the flat bands at the magic angle, and when doped they host the conventional intraband plasmon\cite{stauber2013optical, lewandowski2019intrinsically}. The remote bands are the minibands at higher energy, and transitions that involve them give rise to interband plasmons\cite{stauber2016quasi, hesp2021observation}. Both types of bare plasmon, and the interband plasmons in particular, can be strongly attenuated by intrinsic and extrinsic scattering\cite{ni2018fundamental, yan2013damping}. We overcome these limitations using hyperbolic plasmon--phonon polaritons (HP$^{3}$), formed through hybridization between graphene plasmons\cite{fei2012gate, chen2012optical} and hyperbolic phonon polaritons in hexagonal boron nitride (hBN)\cite{dai2014tunable, caldwell2014sub, brar2014hybrid, dai2015graphene, basov2016polaritons}. Combining the low-loss propagation of hBN modes with the gate-dependent conductivity of graphene, HP$^{3}$ modes remain observable in regimes where bare graphene plasmons do not propagate. The HP$^{3}$ mode thus acts as an indirect sensor of the electronic collective response, which it encodes in two measurable quantities: the polariton wavelength and its damping.

Here we use HP$^{3}$ near-field microscopy to investigate bilayer graphene ranging from AB stacking to $5.5^\circ$ TBG. Our main observation is that, in large-angle TBG, the HP$^{3}$ wavelength responds strongly to the gate only within a narrow window of carrier density. Two crossovers bound this window. The first occurs at low density and is common to graphene systems\cite{dai2015graphene}. Below it, the graphene plasmon lies within the particle--hole continuum and is Landau damped. Above it, the plasmon exits the continuum and hybridizes strongly with the hBN phonon polariton, so that the HP$^{3}$ wavelength grows rapidly with density. The second crossover closes the window. It is observed only for twist angles above $4^\circ$, shifts systematically with twist angle, and arises from the moiré-modified remote bands. As the chemical potential approaches the upper portion of the central band, hybridization with the remote bands flattens the band dispersion. The Drude response then grows only weakly with density, and the HP$^{3}$ wavelength becomes markedly less sensitive to further gating. Continuum-model calculations reproduce both crossovers and identify the second one as a spectroscopic signature of the moiré potential. The resulting gate response is strongly localized in density: a large peak in the polaritonic tunability $d\lambda/dn$ is confined to a narrow, twist-angle-selected window, enabling total wavelength changes approaching $90\%$ over the experimentally accessible density range. More broadly, these results establish remote-band engineering as a route to positioning and controlling the response window of hybrid polariton modes, providing a design principle for tunable polaritonic materials and reconfigurable material cavities.
\section{Principles of the HP$^{3}$ measurement} 

Our approach translates changes in the dynamical electronic response of TBG into changes in a measurable propagating optical mode. We achieve this using hyperbolic plasmon--phonon polaritons, whose wavelength and damping are sensitive to the optical conductivity of the embedded graphene. Electrostatic gating modifies this conductivity, so that monitoring HP$^{3}$ propagation while continuously tuning the carrier density probes the density evolution of the electronic response at the excitation frequency. 

The devices consist of twisted bilayer graphene encapsulated in hBN and placed above a graphite back gate, allowing electrostatic control of the carrier density. The twist angles are controlled during assembly using a calibrated rotation stage and verified through a combination of transport, AFM, and optical measurements (see Samples and Figs.~S1--3 in the Supplementary Information). TBG supports several plasmon modes. The intraband plasmon of the central bands, which becomes weakly dispersive as these bands flatten towards the magic angle\cite{lewandowski2019intrinsically,papaj2020plasmonic,cavicchi2024theory}, coexists with interband and acoustic plasmons\cite{stauber2013optical,stauber2016quasi,hesp2021observation,cavicchi2024theory}. The spectroscopic signatures of plasmons in TBG devices were also detected. Although the measured spectra exhibited features consistent with the theoretical predictions of Cavicchi \textit{et al.}~\cite{cavicchi2024theory} (see Fig.~S4 in the Supplementary Information), we could not resolve propagating plasmon modes at the corresponding wavelengths. We attribute this limitation to the strong damping of the bare plasmonic modes in TBG. The HP$^{3}$ mode studied below instead involves the intraband plasmon, which hybridizes with the hBN phonon polariton.

Figure~1a illustrates the measurement concept. Far-field mid-infrared radiation is focused onto the sample, while Au disks deposited on the hBN surface act as launchers of HP$^{3}$ modes. The momentum of free-space photons is too small to excite these highly confined polaritons directly. The Au disks supply the missing in-plane momentum, enabling their excitation and propagation through the hBN/TBG/hBN heterostructure\cite{dai2017efficiency}.

Figure~1b illustrates how the electronic response controls the HP$^{3}$ mode. At the charge neutrality point, no interference fringes are observed near the scatterers, even though the excitation frequency lies within the hyperbolic Reststrahlen band of hBN and supports phonon-polariton propagation in regions without graphene.
In the TBG region, vertical interband transitions are allowed when $2|E_F|<\hbar\omega$, causing the graphene sheet to absorb at the probe frequency and act as a loss channel for the polariton\cite{li2008dirac}. Upon electrostatic doping ($n=0.7\times10^{13}$~cm$^{-2}$), these transitions become Pauli blocked and clear interference fringes emerge, as previously observed in monolayer graphene\cite{dai2015graphene}. The appearance of propagating fringes therefore demonstrates that the HP$^{3}$ mode is directly sensitive to the dynamical electronic response of the graphene layers. 

\onecolumngrid
\vspace{-1.0ex}
\begin{center}
\includegraphics[width=0.95\textwidth]{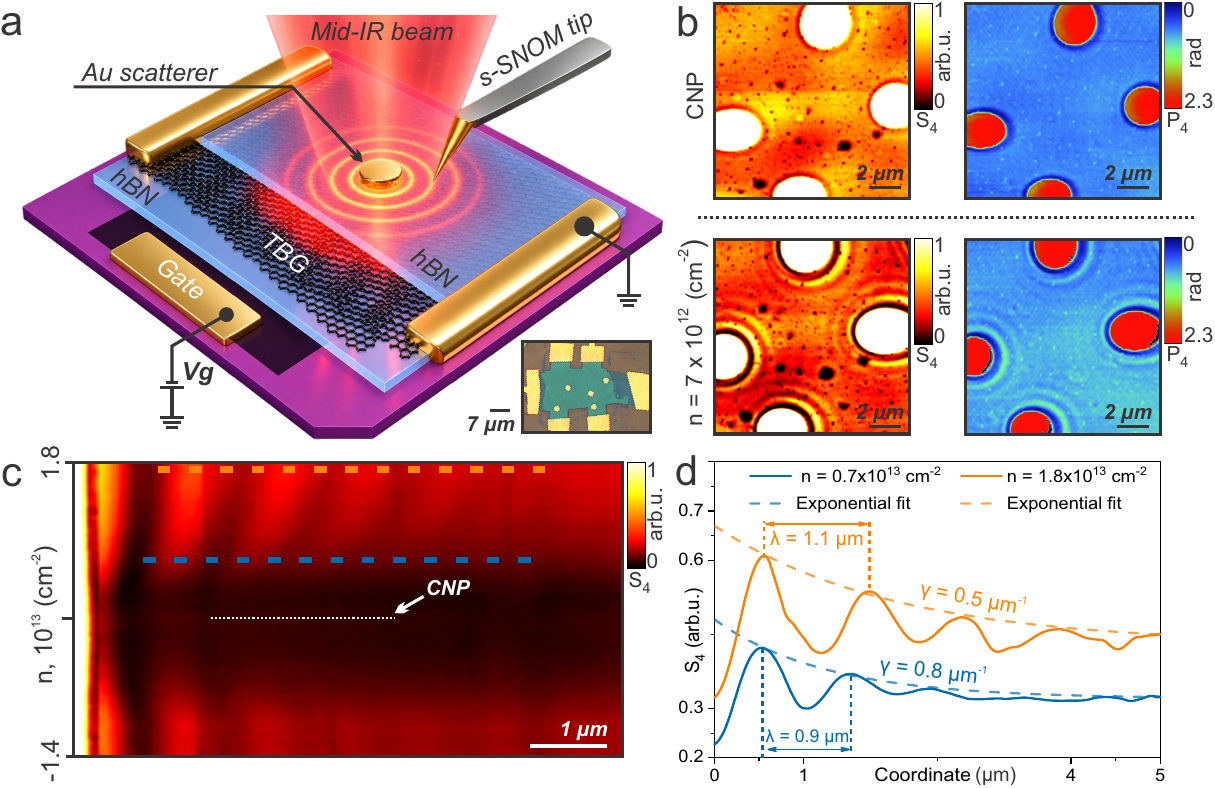}\\[0ex]
\parbox{1\textwidth}{\setlength{\parindent}{0pt}%
\textbf{Figure 1: HP$^{3}$ spectroscopy of twisted bilayer graphene.} \textbf{a}) Schematic of the s-SNOM experiment. TBG encapsulated in hBN is electrostatically doped using a graphite back gate. Au disks launch HP$^{3}$ modes, imaged by the s-SNOM tip. The inset shows an optical image of a representative device. \textbf{b}) Near-field amplitude ($S_{4}$, left) and phase ($P_{4}$, right) maps at 1460~cm$^{-1}$ at the charge neutrality point (CNP, top) and at $n=0.7\times10^{13}$~cm$^{-2}$ (bottom). HP$^{3}$ fringes emerge upon doping as interband absorption becomes Pauli blocked. Bright circular features mark the Au launchers. \textbf{c}) Near-field signal ($S_{3}$) along a line perpendicular to an Au launcher edge as a function of carrier density. Blue and orange lines indicate the profiles in panel d. \textbf{d}) Representative near-field profiles at carrier densities of $0.7\times10^{13}$ and $1.8\times10^{13}$~cm$^{-2}$, with exponential decay fits (see Methods). Fringe spacing and decay envelope yield the HP$^{3}$ wavelength $\lambda$ and damping rate $\gamma$, respectively.}
\end{center}
\vspace{-1.5ex}
\twocolumngrid

To quantify this response, we perform line scans perpendicular to the edge of an Au scatterer, as shown in Fig.~1c. The interference pattern evolves strongly with carrier density: increasing doping produces both longer propagation distances and larger fringe spacing. From representative line profiles in Fig.~1d, we extract the HP$^{3}$ wavelength $\lambda$, determined by the hybridized plasmon--phonon dispersion, and the spatial damping rate $\gamma$, which reflects the dissipation channels available to the collective excitation. In the remainder of the paper, we track $\lambda$ and $\gamma$ as functions of carrier density, twist angle, and temperature. These complementary observables allow us to distinguish changes in the collective-mode dispersion from changes in its propagation losses.

\section{Twist-angle-dependent HP$^{3}$ wavelength and remote-band crossover}
\vspace{-1em}

We now turn to the central question of how the HP$^{3}$ wavelength evolves with carrier density and twist angle. Figure~2a collects the measured wavelength, plotted as its relative change with carrier density (see Eq.~\eqref{1} below), for TBG devices spanning twist angles from $0^\circ$ (AB-stacked bilayer graphene) to $5.5^\circ$ (see Fig.~S5 in the Supplementary Information for the complete set of curves). All data are measured at the same excitation frequency of 1460~cm$^{-1}$ (see Figs.~S6--S8 in the Supplementary Information for additional data, including measurements at 1480~cm$^{-1}$ for the $4.5^\circ$ and $5.5^\circ$ devices). The absolute wavelength differs between devices owing to variations in their geometry, particularly the hBN thicknesses (see Methods). We therefore focus on the evolution of $\lambda$ with carrier density and twist angle rather than on its absolute value.
\onecolumngrid
\vspace{-1.0ex}
\begin{center}
\includegraphics[width=0.95\textwidth]{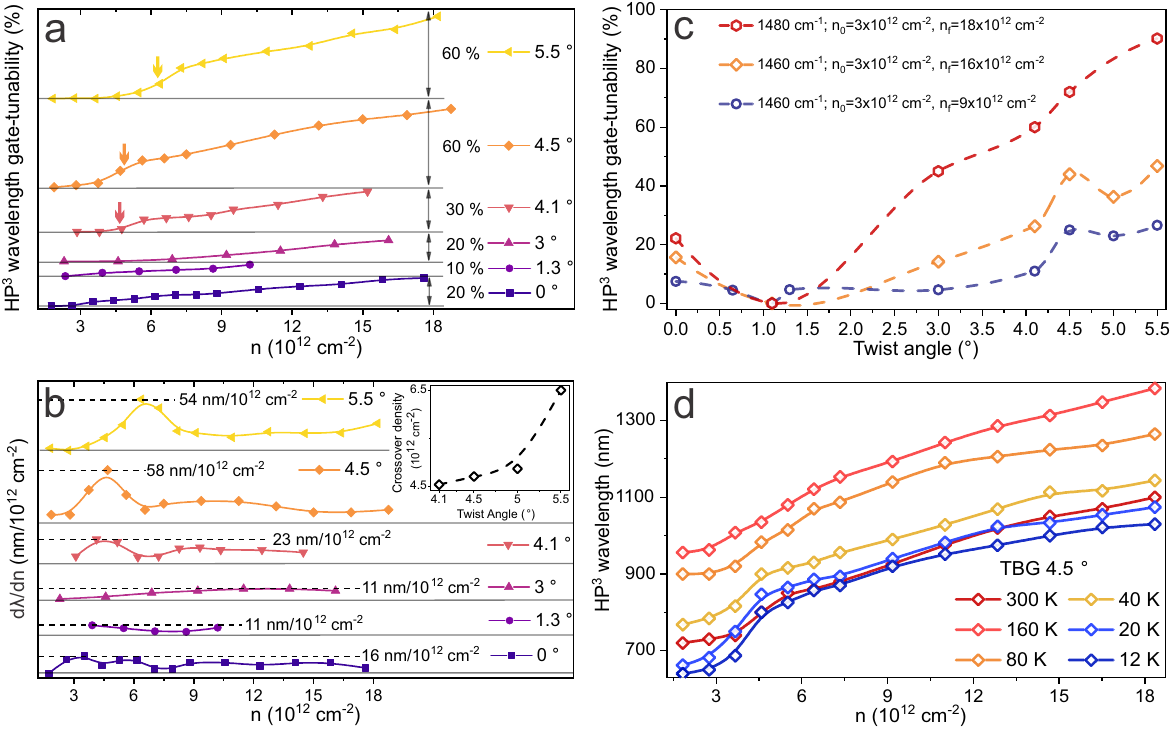}\\[0ex]
\parbox{1\textwidth}{\setlength{\parindent}{0pt}%
\textbf{Figure 2: Twist-angle-dependent HP$^{3}$ crossover and density-localized polaritonic tunability.}
\textbf{a}) HP$^{3}$ wavelength tunability, calculated using Eq.~\eqref{1}, as a function of carrier density for TBG samples with twist angles ranging from $0^\circ$ (AB-stacked bilayer graphene) to $5.5^\circ$. At large twist angles ($\theta \gtrsim 4^\circ$), strong wavelength variation is confined to a narrow density range (indicated by arrows).
\textbf{b}) Wavelength derivative $d\lambda/dn$, showing a maximum in tunability that shifts with twist angle. Inset: crossover density as a function of twist angle.
\textbf{c}) HP$^{3}$ wavelength tunability, calculated using Eq.~\eqref{1}, across different carrier-density ranges and excitation wavenumbers. At 1480~cm$^{-1}$, the wavelength change approaches $90\%$ at the largest twist angles, several times that of the AB-stacked bilayer reference.
\textbf{d}) HP$^{3}$ wavelength as a function of carrier density and temperature for the $4.5^\circ$ device.}
\end{center}
\vspace{-1.0ex}
\twocolumngrid
The AB bilayer graphene sample exhibits the expected monotonic increase of the HP$^{3}$ wavelength with carrier density, consistent with previous studies of graphene-based HP$^{3}$ modes~\cite{dai2015graphene}. The small-angle devices ($0.65^\circ$, $1.3^\circ$, and $3^\circ$) exhibit similar behavior, except for the magic-angle twisted bilayer graphene device, in which HP$^3$ fringes do not emerge at the studied wavenumbers and doping levels and remain strongly damped (see Fig.~S10). Above $4^\circ$, however, the density dependence changes qualitatively: a sharp variation of the wavelength becomes concentrated within a narrow carrier-density window. The emergence of this twist-angle-dependent feature provides the first indication that the HP$^{3}$ mode is sensing an electronic response beyond that of two weakly coupled Dirac cones.

Figure~2b makes this effect explicit by plotting the derivative $d\lambda/dn$. The crossover appears as a distinct maximum whose position shifts to higher carrier density as the twist angle increases (see inset). The systematic angle dependence indicates that the crossover is tied to an electronic scale controlled by the moiré superlattice rather than to a device-specific geometric feature. We will establish its microscopic origin through the continuum-model and electrodynamic calculations presented in the following section.

The crossover also concentrates the gate tunability of the polariton within a narrow density window. Near the crossover density, $d\lambda/dn$ exceeds that of the AB bilayer graphene reference by severalfold. Figure~2c quantifies the total tunability over the experimentally accessible density range as
\vspace{-1.5em}
\begin{equation} 
T=100\%\times\frac{\lambda_f-\lambda_0}{\lambda_0}, 
\label{1}
\end{equation}
\vspace{-1.5em}

\noindent
where $\lambda_0$ and $\lambda_f$ are the HP$^{3}$ wavelengths at the lowest and highest accessible carrier densities, respectively. At an excitation frequency of 1480~cm$^{-1}$, the wavelength change approaches $90\%$ for the largest twist angles, several times the value obtained in the AB bilayer graphene reference. The key distinction is therefore not only a large integrated wavelength change, but a strongly density-localized response whose position is selected by the twist angle.

Figure~2d shows the robustness of the crossover against temperature in the $4.5^\circ$ device. The feature remains visible from 12~K to room temperature, with no clear systematic shift of its characteristic density within the experimental resolution. The overall wavelength scale nevertheless evolves nonmonotonically with temperature. This secondary temperature dependence may reflect changes in the electronic conductivity, the hBN dielectric response, or both. We return to this ambiguity when discussing the HP$^{3}$ damping in Fig.~4.

Taken together, the results of Fig.~2 reveal a twist-angle-dependent spectroscopic signature that emerges only in large-angle TBG and is accompanied by a strong enhancement of polaritonic tunability. Its systematic displacement with twist angle, persistence to room temperature, and absence in the AB bilayer reference point toward an origin in the underlying moiré-modified electronic structure. In the following section, we show that the high-density crossover arises from hybridization between the central and remote moiré bands, which suppresses the density dependence of the plasmonic response and confines the largest HP$^{3}$ tunability to a narrow, twist-angle-selected density window.

\vspace{-1em}
\section{Remote-band hybridization and the origin of the HP$^{3}$ crossover}
\vspace{-1em}

To connect the observed HP$^{3}$ wavelength to the underlying moiré electronic structure, we calculate the electrodynamic response of the full hBN/TBG/hBN heterostructure shown in Fig.~1. The calculation yields the complex $p$-polarized reflection coefficient $r_p(q,\omega)$ at the upper vacuum--hBN interface, where $q$ is the in-plane momentum. The poles of $r_p(q,\omega)$ determine the collective modes of the heterostructure, while maxima of the loss function $\operatorname{Im}r_p(q,\omega)$ provide a convenient numerical measure of their dispersion. At a fixed probe frequency, the corresponding polariton wavelength is $\lambda=2\pi/q$.

The central electronic input to this calculation is the optical conductivity of TBG, which determines the current and surface charge induced at the hBN/TBG interface. We calculate the conductivity using the Bistritzer-MacDonald continuum model\cite{lopes2007graphene, bistritzer2011moire} (see Methods) and decompose it as
\vspace{-0.5em}
\begin{equation} \sigma(\omega) = \sigma_{\rm intra}(\omega) + \sigma_{\rm inter}(\omega). \end{equation}\vspace{-2em}

\noindent The intraband term describes the response of carriers near the Fermi surface and controls the Drude weight of the graphene plasmon. The interband term accounts for transitions between moiré minibands and provides absorptive channels when such transitions are allowed at the probe frequency\cite{stauber2013optical, moon2013optical}. Incorporating both contributions into the electromagnetic boundary conditions yields the HP$^{3}$ response of the complete multilayer structure. A description in terms of the sheet conductivity alone is justified by the long wavelength of the polariton. The momenta probed here, $q\sim10^{7}$~m$^{-1}$, are two orders of magnitude smaller than the moiré reciprocal lattice vectors, $|\mathbf{G}|\gtrsim2\times10^{9}$~m$^{-1}$ at these twist angles. They are also small compared with the Fermi momentum and satisfy $\hbar v_F q\ll\hbar\omega$. The HP$^{3}$ mode therefore averages over many moiré unit cells. For this reason we neglect crystalline local-field corrections\cite{adler1962quantum, wiser1963dielectric}, which couple $q$ to $q+\mathbf{G}$, and the electronic response is close to its local limit, $q\to0$. The weak residual $q$ dependence is retained in the numerical calculation (see Methods).

The moiré potential plays two distinct roles. It defines the moiré Brillouin zone, into which the Dirac cones of the two layers are folded, and it hybridizes the folded bands wherever they cross. Folding alone does not change the long-wavelength response of the two layers. To isolate the role of hybridization, we therefore compare the full calculation with a reference in which the interlayer tunneling amplitudes are set to zero. The reference retains the folded Dirac cones but removes their hybridization. We refer to the two calculations as hybridized and decoupled.

Figure~3a shows the calculated loss function $\operatorname{Im}r_p(q,\omega)$ at representative carrier densities, for the hybridized (top row) and decoupled (bottom row) calculations. The evolution of the spectra can be understood from the TBG band structure shown in Fig.~3b, which identifies four density regimes relevant to the HP$^{3}$ response. At the lowest densities, denoted regime~1, the TBG plasmon overlaps the particle--hole continuum of the central bands and is strongly Landau damped\cite{hwang2007dielectric}. A sharp electronic collective mode therefore does not propagate at the probe frequency. Upon increasing the carrier density into regime~2, vertical interband absorption between the central bands becomes Pauli blocked once $2|E_F|\gtrsim\hbar\omega$. A sharp plasmon then emerges from the particle--hole continuum, and its frequency increases rapidly with density, allowing it to hybridize strongly with the hBN phonon polariton. This generic crossover from an absorptive to a Pauli-blocked response is also present in monolayer and AB bilayer graphene~\cite{li2008dirac, dai2015graphene}, and it occurs in both the hybridized and the decoupled calculation. The behavior specific to the moiré electronic structure appears upon entering regime~3, at a carrier density that we denote $n_{*}$. As the chemical potential approaches portions of the central band that are flattened through hybridization with the remote bands\cite{lopes2007graphene, li2010observation}, the effective Fermi velocity $v_F(n)$ decreases with increasing density. Correspondingly, the Drude weight, which scales schematically as 
\begin{equation} \mathcal{D}(n)\propto v_F(n)\sqrt{n}, \end{equation} 
develops a progressively weaker density dependence than in an isolated Dirac cone\cite{stauber2013optical}. The plasmon remains well defined, but its frequency becomes comparatively insensitive to further gating. Finally, in regime~4, transitions between moiré minibands provide additional absorptive channels and increase the plasmon damping\cite{moon2013optical, yu2019gate}. For the devices studied here, this final regime lies beyond the experimentally accessible density range.

\onecolumngrid
\vspace{-1ex}
\begin{center}
\includegraphics[width=0.90\textwidth]{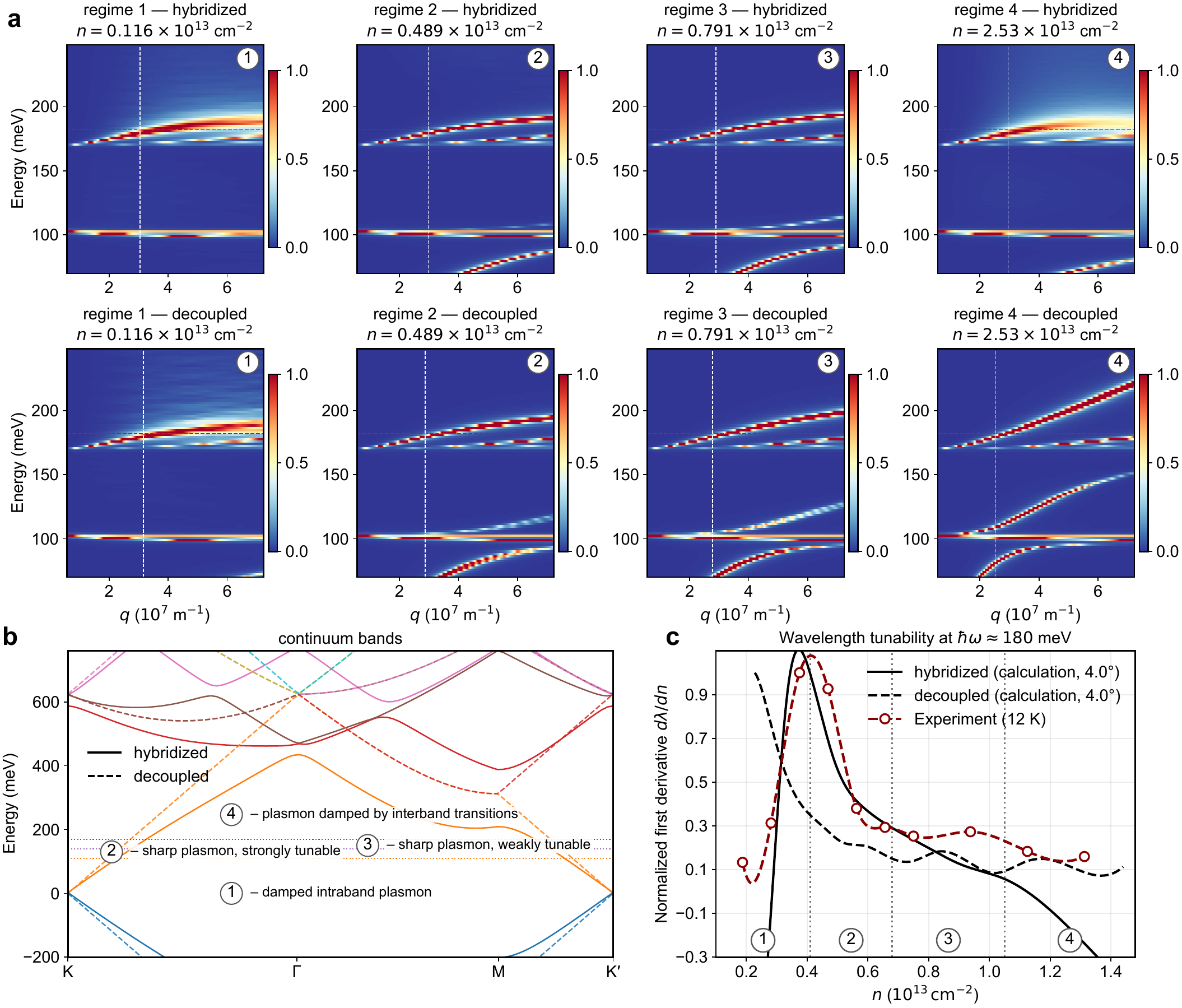}\\[0ex]
\parbox{1\textwidth}{\setlength{\parindent}{0pt}%
\textbf{Figure 3: Remote-band hybridization and the origin of the HP$^{3}$ crossover.} \textbf{a}) Calculated loss function $\textrm{Im}~r_p(q,\omega)$ of the hBN/TBG/hBN heterostructure versus in-plane momentum and frequency at four representative carrier densities, for the hybridized (top row) and decoupled (bottom row) calculations. The horizontal and vertical dashed lines mark the experimental probe frequency and the momentum of the loss maximum at that frequency, respectively. \textbf{b}) Continuum-model band structure of twisted bilayer graphene at $\theta=4.0^\circ$ for the hybridized (solid lines) and decoupled (dashed lines) calculations. Horizontal dotted lines delimit the four carrier-density regimes discussed in the text. \textbf{c}) Normalized derivative $d\lambda/dn$ of the calculated HP$^{3}$ wavelength at the probe frequency versus carrier density for the hybridized (solid line) and decoupled (dashed line) calculations, compared with the measurement on the $4.5^\circ$ device at 12~K (symbols). Each curve is normalized separately. Vertical dotted lines mark the regime boundaries of panel b.} 
\end{center}
\vspace{-1ex}
\twocolumngrid

To summarize, the comparison between the two rows of Fig.~3a isolates the effects of band hybridization. Regimes~1 and 2 have the same character in both calculations. At low density, both calculations exhibit strong electronic damping, while in regime~2 a sharp plasmon emerges and hybridizes with the hBN phonon polariton. Hybridization changes this first crossover only quantitatively. It reduces the Dirac velocity and thereby increases the density of states\cite{lopes2007graphene, luican2011single}, so that the Pauli-blocking condition is reached at a higher carrier density than for decoupled layers. The calculations diverge qualitatively at higher density: with hybridization present, the density evolution of the mode is suppressed in regime~3 and additional damping develops in regime~4. These features are absent when interlayer tunneling is removed, demonstrating that regimes~3 and 4, and with them the boundary $n_{*}$ between regimes~2 and 3, arise from moiré-band hybridization rather than from the generic response of doped Dirac cones.
Figure~3c makes this behavior experimentally explicit by plotting the wavelength derivative, $d\lambda_{\mathrm{HP}^{3}}/dn$, at the probe frequency as a function of carrier density. The absolute wavelength is highly sensitive to the relative thicknesses of the individual layers in the heterostructure. We therefore compare the normalized derivatives to assess the overall trends of the calculated curves against the experimental data (see Methods). In the hybridized calculation, the wavelength varies rapidly in regime~2 but becomes comparatively insensitive to carrier density in regime~3. Regime~2 therefore defines a narrow density window in which the HP$^{3}$ response exhibits the strongest gate tunability, while its high-density boundary, $n_{*}$, marks the onset of the moir\'e-modified response. Notably, the hybridized calculation predicts a pronounced peak in $d\lambda_{\mathrm{HP}^{3}}/dn$, whereas the decoupled calculation exhibits a monotonically decreasing response with increasing carrier density, because its first crossover lies at lower density. The experimental data agree well with the hybridized calculation, further supporting the important role of moir\'e-band hybridization in shaping the observed polaritonic response.

The evolution of the hybrid mode can be understood more directly using a minimal coupled-oscillator model. We represent the hBN phonon polariton and the TBG plasmon as two damped modes coupled with strength $g$, 
\begin{equation} 
\label{eq:oscillators} 
H= \begin{pmatrix} \widetilde{\omega}_{\rm BN} & g\\ g & \widetilde{\omega}_{\rm pl} \end{pmatrix}, \qquad \widetilde{\omega}_{j} \equiv \omega_j-i\gamma_j/2. \end{equation} 
Here, $\omega_{\rm BN}(q)$ and $\gamma_{\rm BN}$ describe the bare hBN phonon-polariton mode, while $\omega_{\rm pl}(q,n)$ and $\gamma_{\rm pl}(q,n)$ describe the bare TBG plasmon. The effective coupling $g(q)$ is set by the electromagnetic mode overlap and therefore depends on the dielectric environment and device geometry. The two complex hybrid-mode frequencies, $\widetilde{\omega}_{\pm}\equiv\Omega_{\pm}-i\Gamma_{\pm}/2$, are the eigenvalues of $H$,
\begin{equation} \label{eq:hybrid_frequencies} \widetilde{\omega}_{\pm} = \frac{ \widetilde{\omega}_{\rm BN} + \widetilde{\omega}_{\rm pl} }{2} \pm \sqrt{ g^2+ \frac{ \left( \widetilde{\omega}_{\rm BN} - \widetilde{\omega}_{\rm pl} \right)^2 }{4} }. \end{equation}
The damping rates enter only through the complex bare frequencies, with $\widetilde{\omega}_{\rm BN}-\widetilde{\omega}_{\rm pl}=(\omega_{\rm BN}-\omega_{\rm pl})-i(\gamma_{\rm BN}-\gamma_{\rm pl})/2$. The square root is therefore itself complex. Although phenomenological, each parameter in Eq.~\eqref{eq:oscillators} has a direct microscopic counterpart. The hBN mode frequency and damping follow from the frequency- and temperature-dependent dielectric tensor of hBN. The TBG plasmon frequency and damping follow from the complex conductivity calculated from the continuum-model bands. The effective coupling $g$ emerges from the electromagnetic boundary-value problem for the complete heterostructure. The oscillator model is therefore not an independent microscopic calculation, but a compact representation of the mode hybridization contained in the full electrodynamic response.

To analyze the role of damping, we define the frequency detuning $\Delta\equiv\omega_{\rm BN}-\omega_{\rm pl}$ and the linewidth difference $\delta\gamma\equiv\gamma_{\rm pl}-\gamma_{\rm BN}$. When the complex detuning is large compared with the coupling, 
\begin{equation} 
g\ll \sqrt{ \Delta^2+ \left(\frac{\delta\gamma}{2}\right)^2 }, 
\end{equation} 
the complex frequency of the predominantly hBN-like branch is, to leading order,  \begin{equation} 
\label{eq:hbn_like_mode} 
\widetilde{\omega}_{\rm hBN-like} \simeq \widetilde{\omega}_{\rm BN} + \frac{g^2}{ \widetilde{\omega}_{\rm BN} - \widetilde{\omega}_{\rm pl} }. 
\end{equation} 
Writing this branch as $\widetilde{\omega}_{\rm hBN-like}\equiv(\omega_{\rm BN}+\delta\Omega_{\rm BN})-i(\gamma_{\rm BN}+\delta\Gamma_{\rm BN})/2$, the corrections to the bare hBN frequency and linewidth are therefore
\begin{align} 
\delta\Omega_{\rm BN} \simeq \frac{ g^2\Delta }{ \Delta^2+\delta\gamma^2/4 }, \quad \delta\Gamma_{\rm BN} \simeq \frac{ g^2\delta\gamma }{ \Delta^2+\delta\gamma^2/4 }. \end{align} 
These expressions show that the observed mode can become predominantly hBN-like in two physically distinct limits. Both correspond to a linewidth mismatch that exceeds the detuning, $\delta\gamma\gg|\Delta|$, for which $\delta\Omega_{\rm BN}\simeq4g^2\Delta/\delta\gamma^2$ and $\delta\Gamma_{\rm BN}\simeq4g^2/\delta\gamma$. The frequency shift is then smaller than the induced broadening by the factor $\Delta/\delta\gamma$. In regime~1, the graphene plasmon is strongly damped by the particle--hole continuum. The large linewidth mismatch suppresses the frequency shift and primarily broadens the hBN-like mode. In regime~4, transitions between moiré minibands again increase $\gamma_{\rm pl}$, leading to a similar suppression of the frequency response. The strongly damped plasmon-like branch is not resolved in either limit.

In regimes~2 and 3, both bare modes are comparatively long lived. Near resonance, where $|\Delta|\lesssim g$ and $g\gg\gamma_{\rm pl},\gamma_{\rm BN}$, the hybrid modes satisfy 
\begin{equation} \Omega_\pm\simeq\omega_0\pm g, \qquad \Gamma_\pm\simeq \frac{ \gamma_{\rm BN}+\gamma_{\rm pl} }{2},
\end{equation} 
with $\omega_0\simeq\omega_{\rm BN}\simeq\omega_{\rm pl}$. In regime~2, $\omega_{\rm pl}(q,n)$ varies rapidly with density, and this density dependence is transferred to the hybrid modes, producing large HP$^{3}$ tunability. In regime~3, the plasmon remains long lived and hybridized with the hBN mode, but its frequency varies only weakly with density because the moiré-modified Drude weight has flattened. The HP$^{3}$ mode consequently remains sharp while becoming much less sensitive to gate voltage.

This theory therefore distinguishes two boundaries of the strongly tunable density window. Its low-density boundary, between regimes~1 and 2, is set by the suppression of Landau damping and the emergence of a coherent graphene plasmon, a feature common to graphene-based systems. Its high-density boundary $n_{*}$, between regimes~2 and 3, marks the onset of the moiré-modified Drude response produced by hybridization between the central and remote bands. The band flattening responsible for this onset originates from the crossing of the Dirac cones of the two layers, whose energy increases with twist angle\cite{li2010observation}. Accordingly, $n_{*}$ is expected to move to higher density as the twist angle increases. Conversely, at smaller twist angles $n_{*}$ approaches the low-density boundary, and no window of coherent, strongly tunable response is expected to remain at the probe frequency, consistent with the absence of the feature below $4^\circ$. It is this second, twist-angle-dependent boundary that provides the spectroscopic signature of remote-band physics and confines the largest HP$^{3}$ gate response to a narrow carrier-density range.

\vspace{-1em}
\section{Temperature- and density-dependent HP$^{3}$ damping} 
\vspace{-1em}

In addition to its wavelength, the spatial damping rate $\gamma$ of the HP$^{3}$ mode provides information about the losses governing polariton propagation. In graphene-based systems, these losses can arise from impurity scattering, electron--phonon interactions, and Landau damping into the electron--hole continuum~\cite{yan2013damping, principi2014plasmon, kaasbjerg2012unraveling, ni2018fundamental}. These mechanisms depend differently on carrier density and temperature, so that measurements of $\gamma$ can provide information complementary to the dispersion studied above.

Figure~4a shows the damping rate $\gamma$ of the $4.5^\circ$ device as a function of probe frequency and carrier density, with the response of a bare hBN slab included as a reference. The damping decreases as the carrier density is increased and then approximately saturates, with the reduction most pronounced at shorter wavelengths. This trend is consistent with the suppression of electronic absorption as the plasmon exits the particle--hole continuum. In the coupled-oscillator picture of the previous section, regimes~2 and 3 contain a comparatively long-lived graphene plasmon that hybridizes with the hBN phonon polariton. The resulting mode can therefore propagate with lower losses than in the low-density, Landau-damped regime. We observe no clear high-density increase in $\gamma$ over the experimentally accessible range, consistent with the theoretical prediction that the strongly damped regime~4 occurs only at higher carrier densities.

Figure~4b shows the density dependence of $\gamma$ at a fixed excitation wavenumber of 1460~cm$^{-1}$ for temperatures between 12 and 300~K (see the complete set of data for $\gamma$ and $\lambda_{HP3}$ in Fig.~S9 of the Supplementary Information). At the lowest and highest temperatures, the damping varies approximately monotonically with density. At intermediate temperatures, additional non-monotonic structure appears, although the experimental uncertainty prevents a detailed assignment of these features. The enhanced damping at low density persists throughout the measured temperature range and broadens upon warming, consistent with thermal smearing of the boundary of the Landau-damped regime.

Although we cannot establish a microscopic origin for the non-monotonic temperature dependence, we note an intriguing correspondence with the temperature-dependent acoustic-phonon properties of graphene. Previous studies have shown that the temperature dependence of phonon-polariton damping in polar crystals, including hBN, is strongly influenced by scattering involving thermally populated acoustic phonons~\cite{ni2021long}. Similarly, temperature-dependent studies of graphene plasmons have identified electron--phonon scattering as one of the relevant dissipation channels, although its contribution can be relatively weak compared with dielectric losses~\cite{ni2018fundamental}.

\onecolumngrid
\vspace{3ex}
\begin{center}
\includegraphics[width=1\textwidth]{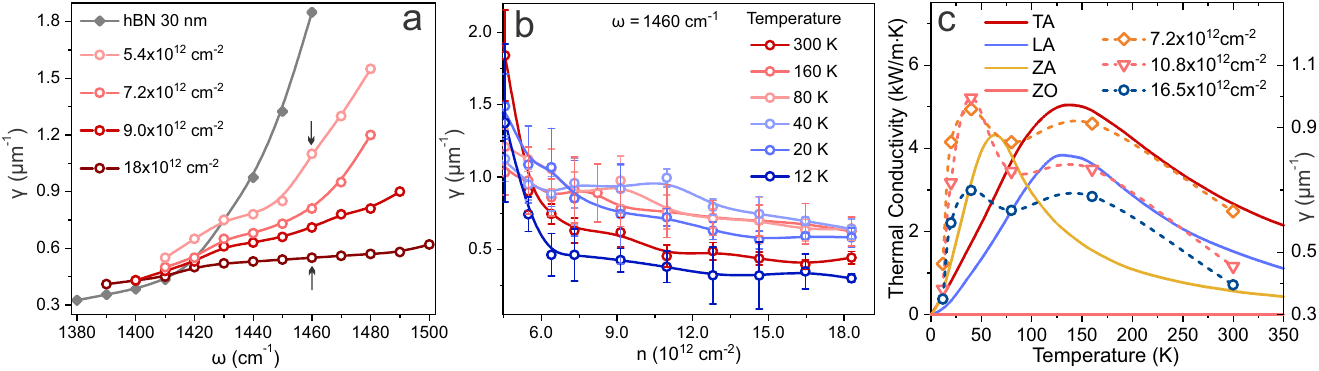}\\[1ex]
\parbox{1\textwidth}{\setlength{\parindent}{0pt}%
\textbf{Figure 4: Density- and temperature-dependent HP$^{3}$ propagation losses.} \textbf{a}) Spatial damping rate $\gamma$ extracted for the $4.5^\circ$ TBG device as a function of excitation wavenumber and carrier density. The grey curve shows the damping rate of a 30~nm-thick bare hBN slab, while the red curves correspond to different carrier densities. Increasing the carrier density suppresses the HP$^{3}$ damping, particularly at higher excitation wavenumbers. \textbf{b}) Carrier-density dependence of $\gamma$ at a fixed excitation wavenumber of 1460~cm$^{-1}$ for temperatures from 12 to 300~K. The enhanced low-density damping persists across the measured temperature range, while additional non-monotonic variations appear at intermediate temperatures. \textbf{c}) Comparison of different phonon contributions to the thermal conductivity of graphene\cite{aksamija2011lattice} (solid lines) with experimentally extracted damping rates $\gamma$ (dashed lines with open symbols), derived from the measurements shown in panel b for several charge carrier densities. }
\end{center}
\twocolumngrid

In our measurements (Fig.~4c), the damping at the lowest ($12$~K) and highest ($300$~K) temperatures follows a relatively smooth trend, whereas an additional enhancement emerges at intermediate temperatures ($40$--$160$~K), exceeding the experimental uncertainty. Interestingly, this temperature window overlaps with the range in which acoustic phonons make substantial contributions to the lattice thermal conductivity of graphene\cite{nika2012two, nika2017phonons}. More specifically, comparing the temperature dependence of the measured HP$^{3}$ damping with the calculated branch-resolved thermal conductivity~\cite{aksamija2011lattice} reveals a suggestive correspondence with the contributions of the ZA and TA (or LA) branches, rather than with the broad total thermal-conductivity profile.

This observation is particularly intriguing in light of recent quantum twisting microscopy\cite{inbar2023quantum} measurements of phonons in twisted graphene systems by Birkbeck \textit{et al.}~\cite{birkbeck2025quantum}, which identified pronounced electron--phonon coupling to the ZA and TA branches, while finding the coupling to the LA branch to be strongly suppressed. Although the physical quantities and experimental conditions are different, the correspondence raises the possibility that the intermediate-temperature enhancement of HP$^{3}$ damping reflects a contribution from acoustic-phonon-mediated scattering, potentially modified by moir\'e-band hybridization. We emphasize that this interpretation remains speculative: the observed temperature dependence does not by itself establish mode-selective electron--phonon coupling, nor does the comparison with lattice thermal conductivity identify the relevant scattering channels. Nevertheless, it motivates further investigation of the interplay between moir\'e electronic states, acoustic phonons, and mid-infrared polariton propagation in twisted graphene systems.

\section{Conclusions}
We used hyperbolic plasmon--phonon polaritons to probe the dynamical electronic response of bilayer graphene from AB stacking to $5.5^\circ$ TBG. At low carrier density, increasing doping suppresses Landau damping and allows a coherent graphene plasmon to hybridize with the hBN phonon polariton. At higher density, large-angle TBG exhibits an additional crossover whose position shifts systematically with twist angle. Continuum-model and electrodynamic calculations trace this crossover to hybridization between the central and remote moiré bands, which weakens the density dependence of the Drude response and, consequently, of the HP$^{3}$ wavelength. HP$^{3}$ spectroscopy therefore provides access to remote-band physics that is difficult to resolve through low-energy transport.

More broadly, our results show that the consequences of moiré band engineering extend well beyond the flat-band and strongly correlated regimes. Far from the magic angle, the low-energy states may resemble weakly coupled Dirac cones, yet the remote bands still reshape the finite-frequency conductivity and collective electrodynamics. Large-angle TBG should therefore not be regarded as electronically trivial: its moiré potential remains visible through the density dependence of its collective modes even when fragile many-body phases are absent.

This remote-band response has a direct functional consequence. The onset of the moiré-modified Drude response confines the largest polaritonic gate response to a narrow, twist-angle-selected density window, with total HP$^{3}$ wavelength changes approaching $90\%$ over the experimentally accessible range. Twist angle thus controls not only the magnitude of the optical response, but also the carrier density at which the response is most sensitive to gating. This establishes moiré-band engineering as a route to positioning and controlling the active window of a polaritonic material rather than fixing its optical properties solely during crystal growth and device fabrication.

The ability to tune the wavelength, confinement, and losses of hBN-based polaritons may also be relevant to the emerging use of hyperbolic materials as electromagnetic environments for controlling quantum matter\cite{schlawin2022cavity, herzig2024high}. Recent work has shown that resonant hBN polariton modes can modify the superconducting response of an adjacent material~\cite{keren2026cavity}. Our results do not demonstrate such control directly, but they identify twist angle and carrier density as additional parameters for engineering the constituent modes of a material cavity. Large-angle TBG/hBN heterostructures may therefore provide a route toward actively reconfigurable material cavities in which the electromagnetic environment, and potentially its coupling to electronic phases, can be adjusted in situ.

\section*{\textsf{Methods}}
\small
\subsection*{Device fabrication.}
\vspace{-1em}
Si$^{++}$/SiO$_2$ substrates were first patterned with metal electrodes using standard photolithography and electron-beam evaporation. Photolithography was performed using a SUSS MicroTec MJB4 mask aligner and SPR-700 photoresist. Following development, Ti/Au (10/50 nm) electrodes were deposited using an Angstrom Engineering electron-beam evaporation system.

hBN, graphite, and graphene flakes were mechanically exfoliated onto Si/SiO$_2$ substrates using adhesive tape and identified by optical microscopy based on their optical contrast. Van der Waals heterostructures were assembled using the conventional dry-transfer cut-and-stack technique\cite{saito2021hofstadter}, which provides atomically clean interfaces and precise control over the relative crystallographic orientation of the constituent layers. Twisted bilayer graphene (TBG) was fabricated by the tear-and-stack approach\cite{kim2016van}, with the relative twist angle controlled with a precision of approximately $0.1^\circ$. The assembled heterostructure was subsequently transferred onto the pre-patterned Si$^{++}$/SiO$_2$ substrate using a PDMS/PC-based transfer technique.

Device patterning was performed by electron-beam lithography using a Raith EBPG5200 system and PMMA 950 A4 resist. Following development, the exposed regions were etched using an Oxford Instruments ICP-RIE system with CHF$_3$/O$_2$ plasma (40/6 sccm, 90 mTorr, 60 W). A second electron-beam lithography step defined the contact geometry and enabled electrical connection to the pre-patterned electrodes. Finally, Cr/Au (2.5/30 nm) contacts were deposited by electron-beam evaporation, forming one-dimensional edge contacts to the graphene layers.

\subsection*{Near-field optical measurements.}
\noindent
Near-field optical measurements were performed using a commercial scattering-type scanning near-field optical microscope (s-SNOM, cryo-neaSCOPE, Neaspec/Attocube) operated in pseudo-heterodyne detection mode. This approach provides background-free measurements of the near-field optical amplitude and phase by demodulating the scattered signal at the tip-oscillation harmonics and their sidebands ($\Omega \pm NM$)\cite{ocelic2006pseudoheterodyne, moreno2017phase,keilmann2004near}. PtIr-coated AFM tips (Nanoworld, 23 nm coating thickness) with a resonance frequency of approximately 250 kHz were used for simultaneous topographic and near-field optical imaging.

The AFM tip was illuminated with a tunable mid-infrared quantum cascade laser (MIRcat, DRS Daylight Solutions Inc.) operating over the wavelength range of 5.5--11.5~$\mu$m, with an incident optical power of 1--2 mW. All measurements presented in this work were acquired at the third or fourth harmonic of the tip-oscillation frequency ($S_3$ or $S_4$, respectively), thereby suppressing the far-field background. Under the experimental conditions, this signal is sensitive to the local out-of-plane electric-field component and provides a measure of the spatially resolved near-field intensity, approximately proportional to $|E_z|^2$.
\noindent
\subsection*{TBG continuum model}
\noindent
We describe the electronic structure of TBG using the Bistritzer-MacDonald continuum Hamiltonian~\cite{lopes2007graphene,bistritzer2011moire,Koshino2018continuum} for electronic states near each monolayer graphene valley, indexed by $\xi=\pm1$. In each valley, the Bloch wavefunction has four components, $\psi=(\psi^{A_1},\psi^{B_1},\psi^{A_2},\psi^{B_2})^{\mathsf T}$, corresponding to the $A$ and $B$ sublattices in layers 1 and 2.

The layers are rotated by $\theta_1=-\theta/2$ and $\theta_2=+\theta/2$, where $\theta$ is the twist angle. Accounting for the twist, the Dirac $K$ point of each layer is located at
\begin{equation}
\begin{aligned}
\mathbf K^{(\ell)}_\xi&=-\xi\frac{4\pi}{3a}R(\theta_\ell)\hat{\mathbf x},\\
R(\phi)&=\begin{pmatrix}
\cos\phi&-\sin\phi\\
\sin\phi&\cos\phi
\end{pmatrix},
\end{aligned}
\label{eq:continuum_dirac_points}
\end{equation}
where $a=0.246$~nm is the graphene lattice constant. The Hamiltonian near the $K$ point is given by
\begin{equation}
H_\xi=
\begin{pmatrix}
h_{1\xi}(-i\boldsymbol\nabla)&U_\xi^\dagger(\mathbf r)\\
U_\xi(\mathbf r)&h_{2\xi}(-i\boldsymbol\nabla)
\end{pmatrix},
\label{eq:bm_model}
\end{equation}
with
\begin{equation}
\begin{aligned}
h_{1\xi}(\mathbf p)&=-\hbar v_F
\bigl[R(+\theta/2)(\mathbf p-\mathbf K^{(1)}_\xi)\bigr]
\cdot(\xi\sigma_x,\sigma_y),\\
h_{2\xi}(\mathbf p)&=-\hbar v_F
\bigl[R(-\theta/2)(\mathbf p-\mathbf K^{(2)}_\xi)\bigr]
\cdot(\xi\sigma_x,\sigma_y),
\end{aligned}
\label{eq:continuum_intralayer}
\end{equation}
where $\sigma_x$ and $\sigma_y$ are Pauli matrices and $v_F$ is the monolayer graphene Fermi velocity. 

The twist produces a moir\'e unit cell with lattice vector magnitude $L_M=a/[2\sin(\theta/2)]$. We describe the Brillouin zone with the primitive moir\'e reciprocal vectors
\begin{equation}
\begin{aligned}
\mathbf G_1&=\frac{4\pi}{\sqrt{3}L_M}
\left(-\frac12,-\frac{\sqrt{3}}2\right),\\
\mathbf G_2&=\frac{4\pi}{\sqrt{3}L_M}(1,0).
\end{aligned}
\label{eq:continuum_reciprocal_vectors}
\end{equation}
In terms of the reciprocal vectors, the interlayer coupling is given by
\begin{equation}
\begin{aligned}
U_\xi(\mathbf r)
&=\begin{pmatrix}u&u'\\u'&u\end{pmatrix}\\
&\quad+\begin{pmatrix}
u&u'e^{-i\xi\varphi}\\
u'e^{i\xi\varphi}&u
\end{pmatrix}e^{i\xi\mathbf G_1\cdot\mathbf r}\\
&\quad+\begin{pmatrix}
u&u'e^{i\xi\varphi}\\
u'e^{-i\xi\varphi}&u
\end{pmatrix}e^{i\xi(\mathbf G_1+\mathbf G_2)\cdot\mathbf r},
\end{aligned}
\label{eq:bm_tunnelling}
\end{equation}
with $\varphi=2\pi/3$. The parameters $u$ and $u'$ are the interlayer tunneling amplitudes between AA and AB sublattice alignment sites, respectively. 

We diagonalize $H_\xi$ in a plane-wave basis. For band index $b$ and moir\'e Bloch wavevector $\mathbf k$, each component is expanded as
\begin{equation}
\psi^X_{b\mathbf k\xi}(\mathbf r)=
e^{i\mathbf k_{\rm mic}\cdot\mathbf r}
\sum_{\mathbf G}C^X_{b\mathbf k\xi}(\mathbf G)e^{i\mathbf G\cdot\mathbf r},
\label{eq:continuum_plane_wave_basis}
\end{equation}
where $X=A_1,B_1,A_2,B_2$ and $\mathbf G=m_1\mathbf G_1+m_2\mathbf G_2$ with integer $m_1,m_2$. Here,
\begin{equation}
\begin{aligned}
\mathbf k_{\rm mic}&=\mathbf k+
\frac{\mathbf K^{(1)}_\xi+\mathbf K^{(2)}_\xi}{2}
-\xi\frac{\sqrt{3}}{2}k_\theta\hat{\mathbf x},\\
k_\theta&=\bigl|\mathbf K^{(1)}_\xi-\mathbf K^{(2)}_\xi\bigr|
=\frac{4\pi}{3L_M}.
\end{aligned}
\label{eq:continuum_momentum_origin}
\end{equation}
Retaining 25 reciprocal vectors, with two sublattices in each of the two layers, we obtain a $100\times100$ Hamiltonian per spin and valley. The moir\'e Brillouin zone is sampled on a uniform $150\times150$ grid in our calculation, and we use the parameters $\theta=4.0^\circ$, $\hbar v_F/a=2.1354$~eV ($v_F\simeq7.98\times10^5$~m\,s$^{-1}$), $u=79.7$~meV, and $u'=97.5$~meV~\cite{Koshino2018continuum}. The decoupled reference calculation sets $u=u'=0$.

To explore the optical properties of TBG, we use the Bloch eigenstates $|u_{a,\mathbf k}\rangle$ of the Bistritzer-MacDonald model to compute the optical conductivity 
\begin{equation}
\sigma_L(q,\omega)=\frac{i e^2\omega}{q^2}\Pi(q,\omega),
\label{eq:tbg_sigma}
\end{equation}
where
\begin{equation}
\begin{aligned}
\Pi(\mathbf q,\omega)&=\frac{g}{A_M N_k}
\sum_{\mathbf k}\sum_{a,b\in\mathcal B}
\frac{\Delta f_{ba}\,|M_{ba}(\mathbf k,\mathbf q)|^2}
{\hbar\omega+\Delta E_{ba}+i\eta},\\
\Delta E_{ba}&=E_{b,\mathbf k+\mathbf q}-E_{a,\mathbf k},\\
\Delta f_{ba}&=f(E_{b,\mathbf k+\mathbf q})-f(E_{a,\mathbf k}).
\end{aligned}
\label{eq:tbg_density_response}
\end{equation}
and $M_{ba}(\mathbf k,\mathbf q)=\langle u_{b,\mathbf k+\mathbf q}|u_{a,\mathbf k}\rangle$. 
Here $g=4$ accounts for spin and valley degeneracy, $N_k=150^2$, $A_M=(\sqrt{3}/2)[a/(2\sin(\theta/2))]^2$ is the moir\'e unit-cell area, and $f(E)=[e^{(E-\mu)/(k_B T)}+1]^{-1}$. The response is evaluated at $T=49.15$~K with a broadening of $\eta=1$~meV. 

\subsection*{Electrodynamics of the hBN/TBG/hBN heterostructure}
\noindent
We combine the TBG conductivity with the anisotropic dielectric function of hBN to calculate the complex $p$-polarized reflection coefficient $r_p(q,\omega)$ at the top hBN-vacuum interface. Its resonances describe the hybridization of the TBG plasmons with the hBN phonon polaritons, appearing as large peaks in $\operatorname{Im}r_p$ shown in Fig.~3a~\cite{dai2015graphene}.

The model of the system, from top to bottom, consists of vacuum (indexed by $j = 0$), a 15~nm thick top hBN layer ($j = 1)$, the TBG sheet, a 15~nm thick bottom hBN layer ($j = 2$), and a graphite gate underneath the hBN approximated as a perfect electrical conductor\cite{menabde2022image}. We will use $z$ to denote the coordinate perpendicular to the interfaces, with the upper hBN-vacuum interface at $z=0$, the TBG lying at $z=-d_1$ and the lower hBN-graphite gate interface at $z=-d_2$, where $d_1=15$~nm and $d_2=30$~nm. 

To determine the field in the hBN layers, we first characterize the optical properties of hBN. hBN is a uniaxial dielectric material whose optical axis is perpendicular to its atomic layers\cite{caldwell2014sub}. Taking the hBN layers to be parallel to the heterostructure interfaces, we can write its dielectric tensor as $\hat\varepsilon=\operatorname{diag}(\varepsilon^t,\varepsilon^t,\varepsilon^z)$, where
\begin{equation}
\varepsilon^\alpha_j(\omega)=\varepsilon^\alpha_\infty
\left[1+\frac{(\omega^\alpha_{\rm LO})^2-(\omega^\alpha_{\rm TO})^2}
{(\omega^\alpha_{\rm TO})^2-\omega^2-i\gamma^\alpha\omega}\right].
\label{eq:lorentz}
\end{equation}
Here $\alpha=t,z$, $j = 1,2$, and $(\varepsilon^t_\infty,\varepsilon^z_\infty)=(4.87,2.95)$. We use $(\omega_{\rm TO}^{t},\omega_{\rm LO}^{t})
=(1367~\mathrm{cm}^{-1},1610~\mathrm{cm}^{-1})\times 2\pi c$
for the transverse and longitudinal optical phonon frequencies for the in-plane component and
$(\omega_{\rm TO}^{z},\omega_{\rm LO}^{z})
=(783~\mathrm{cm}^{-1},828~\mathrm{cm}^{-1})\times 2\pi c$
for the out-of-plane component~\cite{Qu2020hbnParameters}, where $c$ denotes the speed of light.
We also use damping constants
$(\gamma^{t},\gamma^{z})
=(5~\mathrm{cm}^{-1},4~\mathrm{cm}^{-1})\times 2\pi c$
\cite{dai2014tunable}.

We write the $p$-polarized electric field in each region as
$\mathbf{E}^{(j)}(x,z,t)=\bigl(E_x^{(j)}(z),0,E_z^{(j)}(z)\bigr)e^{i(qx-\omega t)}$,
where $q$ is the in-plane wave number, $\omega$ is the angular
frequency, 
\begin{equation}
    E_x^{(j)}(z)=A_j e^{-ik_j z}+B_j e^{ik_j z}
\end{equation}
and
\begin{equation}
    E_z^{(j)}(z)=\frac{qQ_j}{\varepsilon_j^z}
\bigl(A_j e^{-ik_j z}-B_j e^{ik_j z}\bigr).
\end{equation}
$A_j$ and $B_j$ are complex amplitudes (for $j=2$ the coordinate $z$ in the exponentials is measured from the TBG plane, $z\to z+d_1$), and
\begin{equation}
k_j=\sqrt{\varepsilon_j^t\left(\frac{\omega^2}{c^2}-\frac{q^2}{\varepsilon_j^z}\right)},
\qquad Q_j=\frac{\varepsilon_j^t}{k_j},
\label{eq:kz}
\end{equation}
with $\varepsilon_0^t=\varepsilon_0^z=1$. The accompanying magnetic field in each layer is given by
$\mathbf{H}^{(j)}(x,z,t)=\bigl(0,H_{y}^{(j)}(z),0\bigr)e^{i(qx-\omega t)}$.  We fix the incident amplitude $A_0=1$, leaving five unknowns: the reflected amplitude $B_0$ and the four amplitudes $A_1,B_1,A_2,B_2$. 

At the top vacuum--hBN interface, the tangential fields are continuous. At the TBG sheet, where a surface current accumulates due to the sheet conductivity $\sigma_L(q,\omega)$, $E_x$ remains continuous, while $H_y^{(2)}-H_y^{(1)}=J_x$ with $J_x=\sigma_L E_x$. The charge density of the sheet $\rho=qJ_x/\omega$ follows from charge conservation. Defining
\begin{equation}
\mathcal S=\frac{\sigma_L(q,\omega)}{\varepsilon_0\omega},
\label{eq:sheet_phases}
\end{equation}
the boundary conditions at each interface give the system of equations
\begin{equation}
\begin{aligned}
1+B_0&=A_1+B_1,\\
Q_0(1-B_0)&=Q_1(A_1-B_1)
\end{aligned}
\end{equation}
at the hBN-vacuum interface ($z = 0$),
\begin{equation}
\begin{aligned}
A_1e^{ik_1d_1}+B_1e^{-ik_1d_1}&=A_2+B_2 \\
Q_1\left(A_1e^{ik_1d_1}-B_1e^{-ik_1d_1}\right)
&=Q_2(A_2-B_2)+\mathcal S(A_2+B_2)\\
\end{aligned}
\end{equation}
at the hBN-TBG interface ($z = -d_1$),
and 
\begin{equation}
A_2e^{ik_2(d_2 - d_1)}+B_2e^{-ik_2(d_2 - d_1)}=0
\end{equation}
at the graphite gate ($z = -d_2$) where $E_x^{(2)}(-d_2) = 0$.

These boundary conditions form a system of equations for the unknown amplitudes, yielding a closed expression for the reflection coefficient
\begin{equation}
r_p(q,\omega)= -\frac{B_0}{A_0} = 
\frac{(Q_1-Q_0)F_+-(Q_1+Q_0)e^{2ik_1d_1}F_-}
{(Q_1+Q_0)F_+-(Q_1-Q_0)e^{2ik_1d_1}F_-}.
\label{eq:rp}
\end{equation}
where
\begin{equation}
\begin{aligned}
F_+&=(Q_1+Q_2+\mathcal S)-e^{2ik_2(d_2-d_1)}(Q_1-Q_2+\mathcal S),\\
F_-&=(Q_1-Q_2-\mathcal S)-e^{2ik_2(d_2-d_1)}(Q_1+Q_2-\mathcal S).
\end{aligned}
\label{eq:reflection_factors}
\end{equation}
To compute the wavelength tunability in Fig.~3c, we use the frequency $\hbar\omega_c=179.8$~meV and compute $r_p(q,\omega_c)$ as a function of $q$ for various chemical potentials. In the low density limit, where the TBG plasmon mode is heavily damped, the profile of $\operatorname{Im}r_p$ along $q$ is noisy. We apply a Gaussian filter with a small standard deviation of  $4.09\times10^5$~m$^{-1}$ and compute the resonance wavelength $\lambda(n)=2\pi/q_{\rm max}$ using the wavevector $q_{\rm max}$ that maximizes $\operatorname{Im}r_p$ after Gaussian filtering. Finally, the dimensionless tunability is computed by
\begin{equation}
\begin{aligned}
\mathcal T(n)&=\frac{\lambda(n)-\lambda(n_{\rm ref})}{\lambda(n_{\rm ref})},\\
n_{\rm ref}&=0.25\times10^{13}~\mathrm{cm}^{-2}.
\end{aligned}
\label{eq:wavelength_tunability}
\end{equation}
Figure~3c shows the derivative of $\mathcal T(n)$ with respect to $n$, which is proportional to $d\lambda/dn$, with each curve normalized separately. The comparison with the measurement in that panel is qualitative, since the calculation is performed for $\theta=4.0^\circ$ and the measurement on the $4.5^\circ$ device. In particular, the calculated derivative changes sign near $n\approx1.1\times10^{13}$~cm$^{-2}$, at the onset of regime~4, whereas the measured derivative remains positive up to the highest density. This difference is consistent with the larger twist angle of the device, for which the remote-band scale moves to higher density.

\bibliographystyle{naturemag-sergi}
\bibliography{HP3.bib}

\vspace*{-0.2cm}
\section*{\textsf{Acknowledgements}}
\vspace*{-0.2cm}
\noindent
P.S. acknowledges support from the US National Science Foundation CAREER award No. 2542595. B.B. acknowledges support from the Gordon and Betty Moore Foundation through Grant GBMF14251 to the American Chemical Society for the ACS-Moore Postdoctoral Bridge Award in Quantum Materials. C.Y. gratefully acknowledges support from the Eddleman Quantum Institute Postdoctoral Fellowship and the Moore Foundation Postdoctoral Fellowship. C. L. was supported in part by start-up funds from Florida State University and the National High Magnetic Field Laboratory and in part by NSF CAREER grant No. DMR-2543710. The National High Magnetic Field Laboratory is supported by the National Science Foundation through NSF/DMR-2128556 and the State of Florida. K.W. and T.T. acknowledge support from the JSPS KAKENHI (Grant Numbers 21H05233 and 23H02052) and World Premier International Research Center Initiative (WPI), MEXT, Japan.

\vspace*{-0.2cm}
\section*{\textsf{Author contributions}}
\vspace*{-0.2cm}
{\small
\noindent
P.S. and B.B conceived and designed the experiment. B.B. and X.Z. fabricated samples and performed the measurements. The data were analyzed and interpreted by B.B., C.Y., C.L. and P.S. C.Y. performed the theoretical calculations with input from C.L. The hBN crystals were provided by K.W. and T.T. B.B., C.Y., C.L. and P.S. wrote the manuscript with input from all authors. P.S. supervised the work.
}

\vspace*{-0.2cm}
\section*{\textsf{Competing Financial Interests}}
\vspace*{-0.2cm}
{\small
\noindent
The authors declare no competing financial interests.
}

\vspace*{-0.2cm}
\section*{\textsf{Data Availability Statement}}
\vspace*{-0.2cm}
{\small
\noindent
The data that support the findings of this study is available from the corresponding authors upon reasonable request.
}
\newpage
\end{document}


\title{\Large\textsf{\papertitle}}

\author{Bogdan Borodin}
\email{bborodin@nd.edu}
\affiliation{\footnotesize Department of Physics and Astronomy, University of Notre Dame, Notre Dame, IN 46556,~USA}
\affiliation{\footnotesize Stavropoulos Center for Complex Quantum Matter, University of Notre Dame, Notre Dame, IN 46556,~USA}

\author{Xuanpu Zhang}
\affiliation{\footnotesize Department of Physics and Astronomy, University of Notre Dame, Notre Dame, IN 46556,~USA}
\affiliation{\footnotesize Stavropoulos Center for Complex Quantum Matter, University of Notre Dame, Notre Dame, IN 46556,~USA}

\author{Christopher Yang}
\affiliation{\footnotesize Caltech}

\author{Kenji Watanabe}
\affiliation{\footnotesize Research Center for Electronic and Optical Materials, National Institute for Materials Science, 1-1 Namiki, Tsukuba 305-0044,~Japan}

\author{Takashi Taniguchi}
\affiliation{\footnotesize Research Center for Materials Nanoarchitectonics, National Institute for Materials Science,  1-1 Namiki, Tsukuba 305-0044,~Japan}

\author{Cyprian Lewandowski}
\email{clewandowski@magnet.fsu.edu}
\affiliation{\footnotesize  Florida State University}

\author{Petr Stepanov}
\email{pstepano@nd.edu}
\affiliation{\footnotesize Department of Physics and Astronomy, University of Notre Dame, Notre Dame, IN 46556,~USA}
\affiliation{\footnotesize Stavropoulos Center for Complex Quantum Matter, University of Notre Dame, Notre Dame, IN 46556,~USA}

\keywords{phonon-polariton, polaritonic optics, strong confinement, hexagonal boron nitride, focusing}


\begin{abstract}
\vspace*{0.2cm}

\end{abstract}

\clearpage
\onecolumngrid

\tableofcontents

\clearpage
\twocolumngrid

\renewcommand{\figurename}{Fig.}
\setcounter{equation}{0}
\setcounter{figure}{0}
\setcounter{table}{0}
\makeatletter
\renewcommand{\theequation}{S\arabic{equation}}
\renewcommand{\thefigure}{S\arabic{figure}}
\renewcommand{\thetable}{\arabic{table}}
\renewcommand{\bibnumfmt}[1]{[#1]}
\renewcommand{\citenumfont}[1]{#1}

\onecolumngrid

\section{Device fabrication and characterization}
\subsection{Large twist-angle device stacking}
\begin{figure}[h!]
    \centering
    \includegraphics[width=0.9\linewidth]{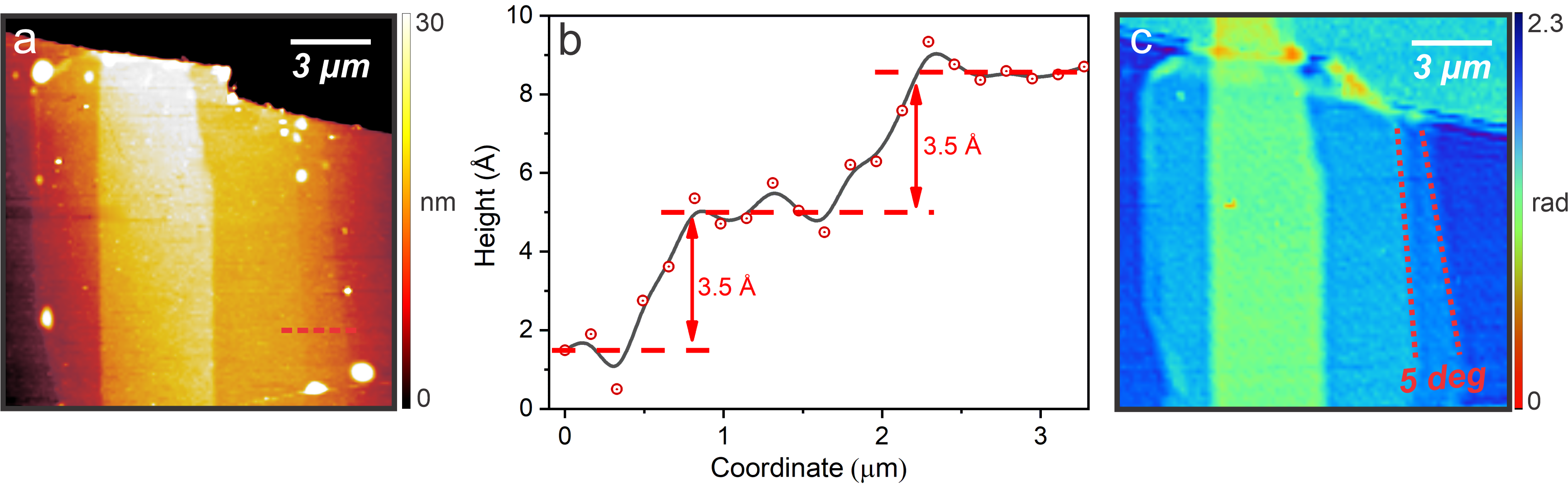}
\caption{Characterization of the $5^\circ$-twist-angle graphene stack. \textbf{a}) AFM topography. \textbf{b}) Height profile along the red line in panel~\textbf{a}. \textbf{c}) Near-field optical phase image acquired at $\omega = 1450~\mathrm{cm}^{-1}$.}
    \label{fig:S1}
\end{figure}

To fabricate large-twist-angle devices, we used a standard ``cut-and-stack'' technique. hBN and graphene flakes were obtained by mechanical exfoliation and transferred onto Si/SiO$_2$ substrates for optical identification. Flakes with the desired thicknesses were selected for device fabrication. A dry-transfer method using PDMS/PC stamps was used to pick up hBN flakes and graphene, which was subsequently cut with an AFM tip. \autoref{fig:S1} shows a $\sim5^\circ$ twisted stack prior to device fabrication.
\vspace{-1em}
\vspace{-1em}
\autoref{fig:S1}a shows the AFM topography of the stacked structure. The bright vertical stripe corresponds to a graphite flake used as the bottom gate. The device area above the graphite gate is free of visible bubbles. Two graphene flakes can be clearly distinguished in the image. \autoref{fig:S1}b shows the height profile along the red line indicated in \autoref{fig:S1}a, revealing two steps corresponding to the monolayer thickness. \autoref{fig:S1}c shows the near-field optical phase image measured at $\omega = 1450~\mathrm{cm}^{-1}$. The near-field image provides substantially higher contrast than the topography, allowing the relative orientation of the two graphene layers to be determined directly.
\subsection{Transport characterization}
\begin{figure}[h!]
    \centering
    \includegraphics[width=0.9\linewidth]{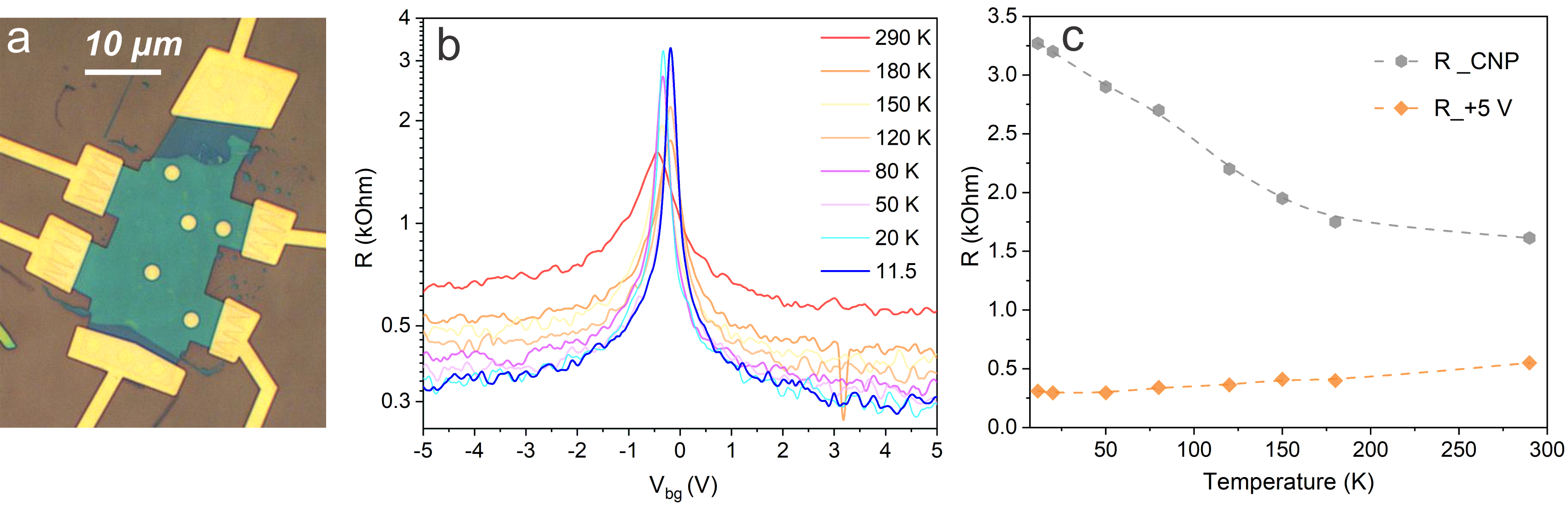}
    \caption{Transport characterization of a $4.5^\circ$ twisted graphene device. \textbf{a}) Optical image of the device. \textbf{b}) Two-terminal resistance as a function of back-gate voltage and temperature. \textbf{c}) Two-terminal resistance as a function of temperature at the CNP and at $V_{\mathrm{bg}}=+5$~V.}
    \label{fig:S2}
\end{figure}

One of the main challenges in studying large-twist-angle graphene samples is reliably determining the twist angle. The carrier density required to observe moir\'e features increases rapidly with twist angle. Consequently, within a practical gate-voltage range, these features become inaccessible, and transport measurements typically reveal only the charge neutrality point (CNP), as shown in \autoref{fig:S2}.
\vspace{-1em}
\vspace{-1em}
\autoref{fig:S2}a shows an optical image of the $4.5^\circ$ twisted device. \autoref{fig:S2}b and \autoref{fig:S2}c show representative two-terminal resistance measurements, revealing the evolution of the CNP from 290~K down to 11.5~K.

\subsection{Gate capacitance calibration}

We determined the gate-voltage-to-carrier-density conversion using TBG devices with twist angles close to the magic angle, where well-resolved Landau levels allow the carrier density to be determined directly. At a fixed magnetic field, the filling factor is given by
\begin{equation}
    \nu = \frac{nh}{eB},
    \label{eq:filling_factor}
\end{equation}
and the fourfold spin and valley degeneracy of the graphene Landau levels gives $\Delta\nu=4$ between neighboring Landau levels. The corresponding carrier-density difference is therefore
\begin{equation}
    \Delta n = \frac{\Delta\nu eB}{h}
    = \frac{4eB}{h}.
    \label{eq:landau_density}
\end{equation}

\begin{figure}[h!]
    \centering
    \includegraphics[width=0.5\linewidth]{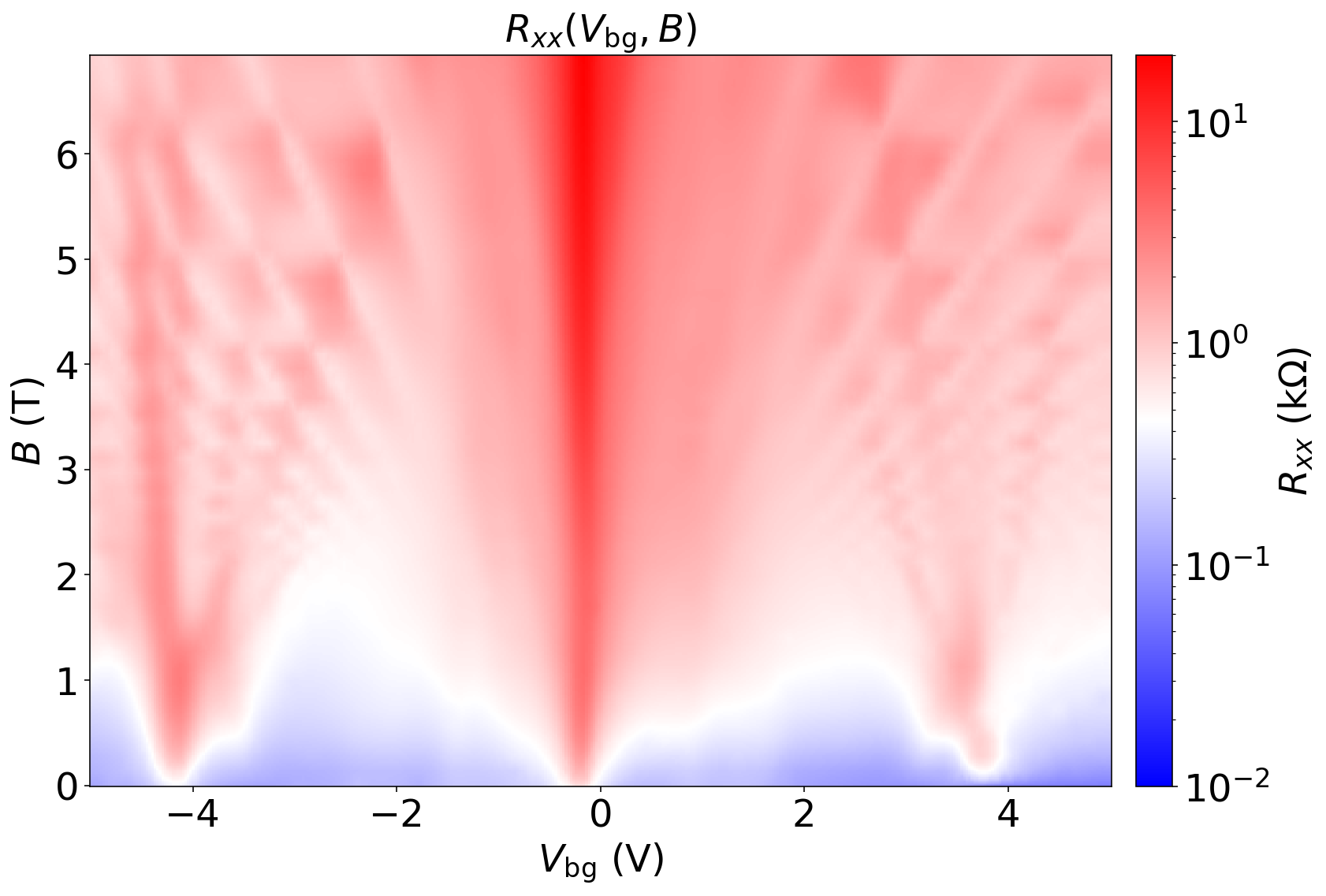}
   \caption{Landau-fan characterization of the $1.3^\circ$ TBG device used for gate-capacitance calibration. Four-terminal longitudinal resistance, $R_{xx}$, as a function of back-gate voltage, $V_{\mathrm{bg}}$, and magnetic field, $B$.}
    \label{fig:S3}
\end{figure}

The Landau fan used for this calibration is shown in \autoref{fig:S3}.

The gate-voltage separation between neighboring Landau levels, $\Delta V_{\mathrm{g}}$, then determines the gate-voltage-to-density conversion coefficient,
\begin{equation}
    \alpha_{\mathrm{g}} = \frac{\Delta n}{\Delta V_{\mathrm{g}}},
    \label{eq:gate_conversion}
\end{equation}
such that
\begin{equation}
    n = \alpha_{\mathrm{g}}\left(V_{\mathrm{g}}-V_{\mathrm{CNP}}\right).
    \label{eq:density_conversion}
\end{equation}

We use this experimentally determined conversion coefficient as a reference for devices with larger twist angles. Because the gate capacitance is primarily determined by the thickness of the hBN dielectric, the coefficient was scaled between devices according to
\begin{equation}
    \alpha_{\mathrm{g}} \propto \frac{1}{d_{\mathrm{hBN}}},
    \label{eq:capacitance_scaling}
\end{equation}
where $d_{\mathrm{hBN}}$ is the thickness of the hBN layer separating the graphene from the gate. Thus, for a device with hBN thickness $d_2$, the conversion coefficient was obtained from the calibrated value $\alpha_1$ of the reference device as
\begin{equation}
    \alpha_2 = \alpha_1\frac{d_1}{d_2}.
    \label{eq:conversion_scaling}
\end{equation}
This procedure accounts for small device-to-device variations in hBN thickness. The resulting conversion coefficients are of order $10^{12}~\mathrm{cm}^{-2}\mathrm{V}^{-1}$ and were used to convert gate voltage to carrier density in the $5^\circ$ devices.

\newpage
\section{Spectroscopy of plasmons in large-angle TBGs}
\begin{figure}[h!]
    \centering
    \includegraphics[width=0.85\linewidth]{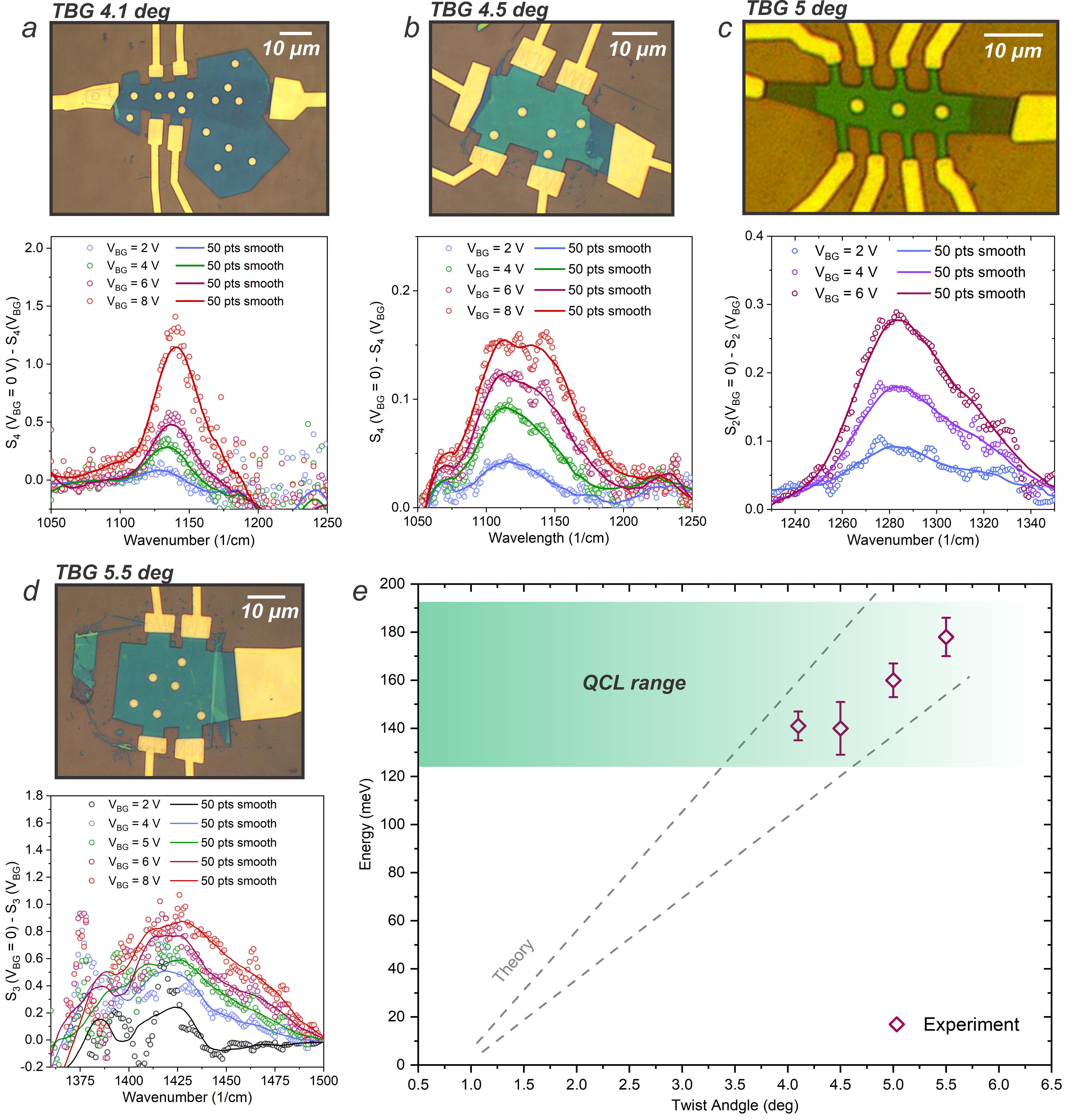}
    \vspace{-1em}
    \caption{Spectroscopic characterization of large-angle TBG devices. \textbf{a--d}) Optical images and background-subtracted near-field optical amplitude spectra acquired at several back-gate voltages for TBG devices with twist angles ranging from $4.1^\circ$ to $5.5^\circ$. The spectrum acquired at zero back-gate voltage served as the background. \textbf{e}) Experimentally measured peak positions (red open diamonds), the theoretically predicted energy range of the COM plasmon as a function of twist angle (grey dashed lines), and the spectral range accessible with our QCL (green area).}
    \label{fig:S4}
\end{figure}

\vspace{-1em}
Several peculiar plasmonic modes have been predicted in TBG by Polini \textit{et al.}~\cite{cavicchi2024theory}. In particular, a center-of-mass (COM) plasmonic mode is predicted to exist over a broad range of twist angles from $1^\circ$ to $6^\circ$, with a substantial portion of its spectral range lying in the mid-infrared and therefore accessible with our QCL laser. According to these predictions, the plasmon energy depends on the twist angle, while its frequency also depends on the carrier density, since a finite carrier density is required to support a plasmonic response. \autoref{fig:S4} shows near-field optical amplitude spectra measured on several TBG devices with twist angles ranging from $4.1^\circ$ to $5.5^\circ$.

\vspace{-1em}
\vspace{-1em}

\autoref{fig:S4}a--d show optical images and background-subtracted near-field optical amplitude spectra measured at several back-gate voltages. The spectrum acquired at zero back-gate voltage served as the background. The observed resonances shift to higher wavenumbers with increasing back-gate voltage, consistent with a gate-dependent plasmonic response. \autoref{fig:S4}e summarizes the measured resonance positions as a function of twist angle. The grey dashed lines indicate the theoretically predicted energy range of the COM plasmon~\cite{cavicchi2024theory}, while the green area denotes the spectral range accessible with our QCL. The red open diamonds show the experimentally measured positions of the gate-dependent resonances, with the error bars determined from the spectral peak widths. The measured gate-dependent features fall within the theoretically predicted COM-plasmon energy range across the investigated twist-angle range, in good agreement with the theoretical predictions.

\newpage
\vspace{-1em}
\section{HP3 wavelength dependence at room temperature for all examined twist angles at $\omega = 1460$ and $1480~\mathrm{cm}^{-1}$}
\vspace{-1em}

\begin{figure}[h!]
    \centering
    \includegraphics[width=0.85\linewidth]{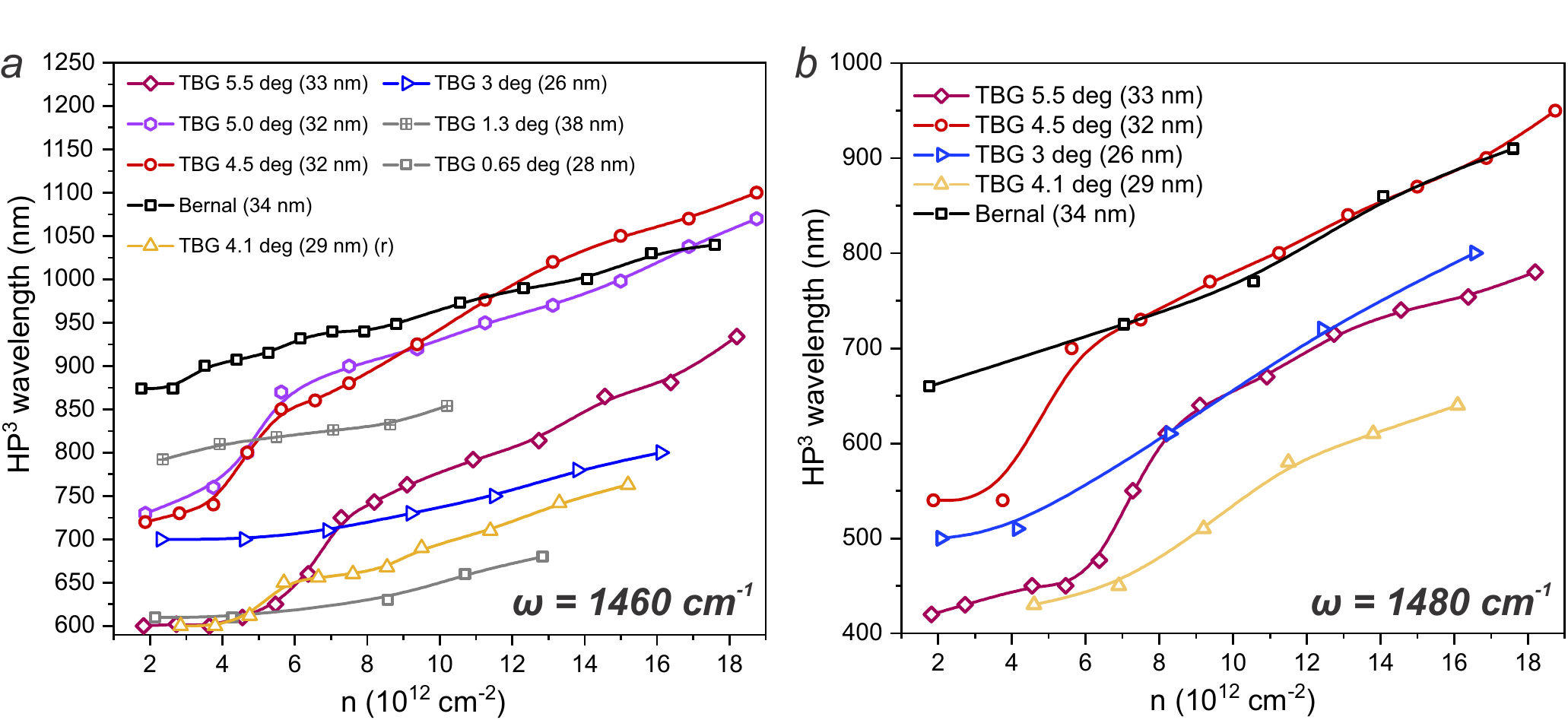}
    \vspace{-1em}
    \caption{\textbf{Gate-dependent HP3 wavelength in TBG samples.}
    \textbf{a}) HP3 wavelength at $\omega = 1460~\mathrm{cm}^{-1}$.
    \textbf{b}) HP3 wavelength at $\omega = 1480~\mathrm{cm}^{-1}$.}
    \vspace{-1em}
    \label{fig:SI5}
\end{figure}

\autoref{fig:SI5} summarizes the HP3 wavelength for all TBG samples studied here. Panels~(a) and (b) show the wavelength at $\omega = 1460~\mathrm{cm}^{-1}$ and $1480~\mathrm{cm}^{-1}$, respectively. Measurements at $\omega = 1480~\mathrm{cm}^{-1}$ were less suitable because of the stronger damping at this higher wavenumber. Therefore, $\omega = 1460~\mathrm{cm}^{-1}$ was selected as the primary wavenumber for the main investigations. 
\vspace{-1em}
\vspace{-1em}

\vspace{-1em}
\subsection{Additional measurements of gate-dependent HP3 in 4.5$^\circ$ and 5.5$^\circ$ samples at room temperature}

\begin{figure}[h!]
    \centering
    \includegraphics[width=0.73\linewidth]{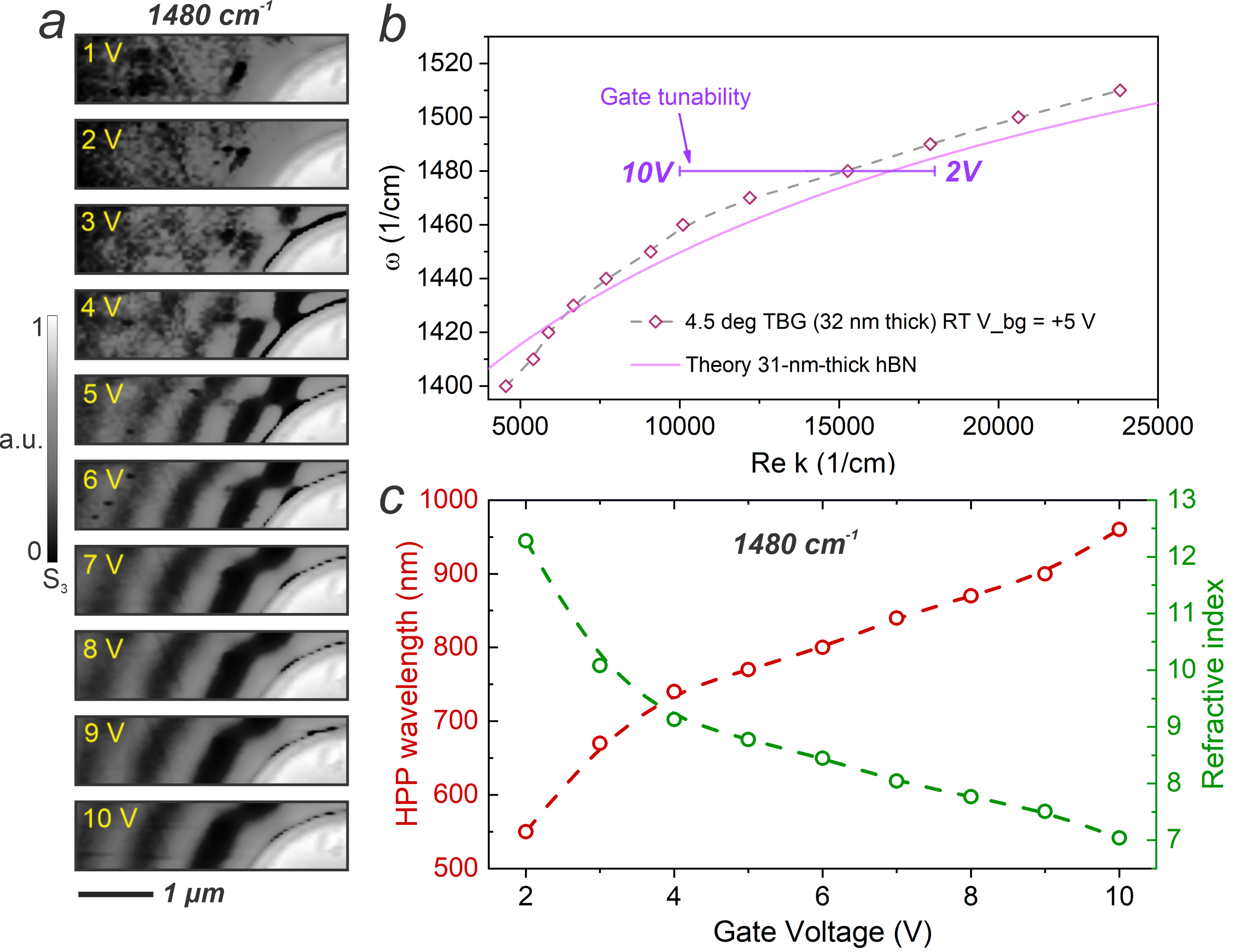}
    \vspace{-1em}
    \caption{\textbf{Gate-dependent HP3 propagation and dispersion in the 4.5$^\circ$ TBG sample.}
    \textbf{a}) Near-field optical amplitude maps measured at 1480~cm$^{-1}$ for gate voltages ranging from 1 to 10~V, showing real-space propagation of HP3 and its gate-dependent evolution.
    \textbf{b}) HP3 dispersion measured at a gate voltage of +5~V.
    \textbf{c}) HP3 wavelength at 1480~cm$^{-1}$ as a function of gate voltage and the corresponding effective refractive index extracted from the measured wavelength.}
    \vspace{-1em}
    \label{fig:SI6}
\end{figure}

\vspace{-1em}
\autoref{fig:SI6} presents additional measurements of gate-dependent HP3 propagation in the $4.5^\circ$ twisted bilayer graphene sample. Panel~(a) shows near-field optical amplitude maps acquired at $1480~\mathrm{cm}^{-1}$ for gate voltages ranging from 1 to 10~V, illustrating the real-space propagation of HP3 and its evolution with gate voltage. Panel~(b) shows the HP3 dispersion measured at a gate voltage of $+5$~V. Panel~(c) summarizes the gate dependence of the HP3 wavelength at $1480~\mathrm{cm}^{-1}$ and the corresponding effective refractive index, extracted from the dispersion and real-space propagation measurements.

\vspace{-2em}
\begin{figure}[H]
    \centering
    \includegraphics[width=0.73\linewidth]{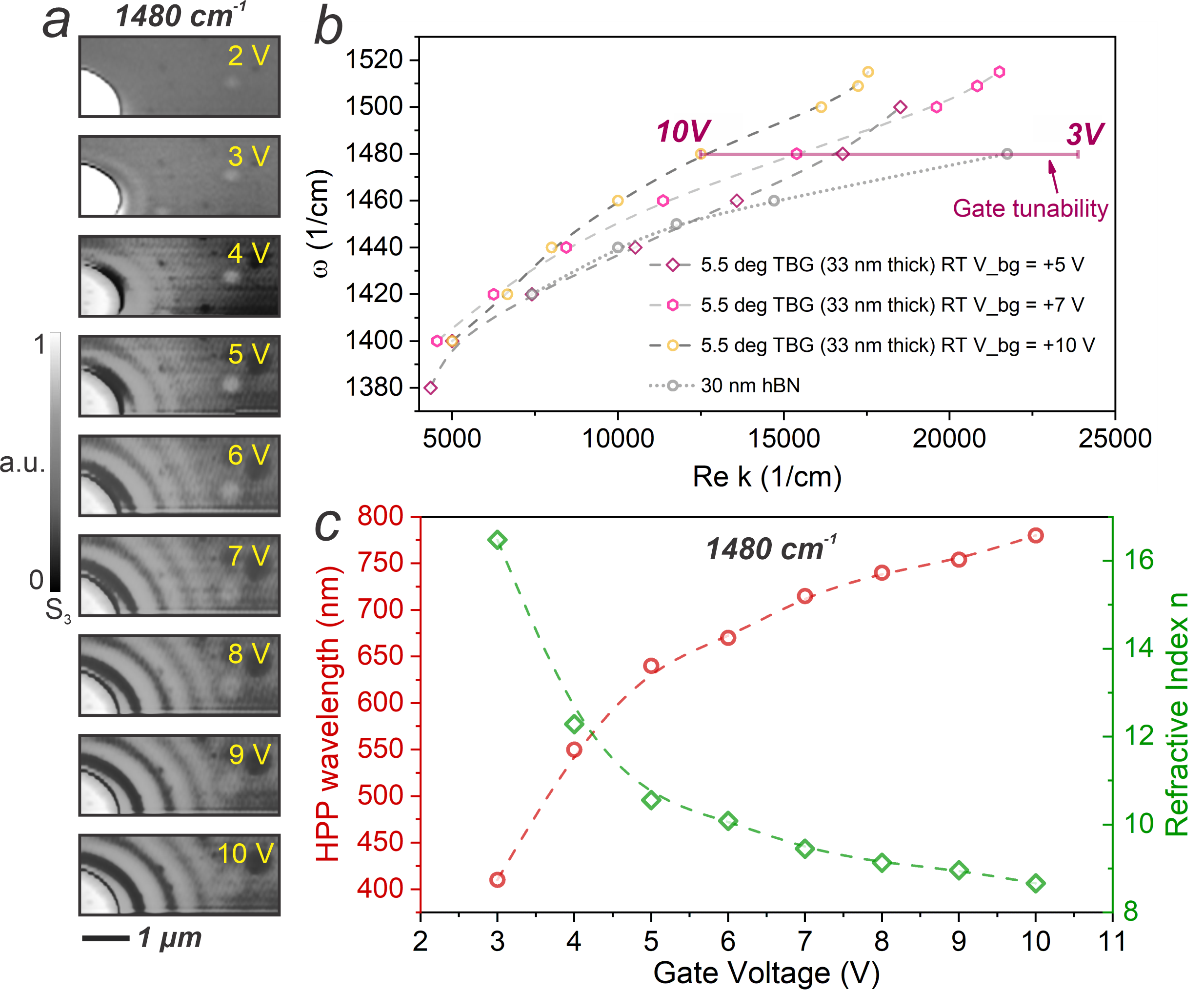}
    \caption{\textbf{Gate-dependent HP3 propagation and dispersion in the 5.5$^\circ$ TBG sample.}
    \textbf{a}) Near-field optical amplitude maps measured at 1480~cm$^{-1}$ for gate voltages ranging from 1 to 10~V, showing real-space HP3 propagation and its evolution with gate voltage.
    \textbf{b}) HP3 dispersion measured at several gate voltages, demonstrating gate-dependent evolution of the polariton dispersion.
    \textbf{c}) HP3 wavelength at 1480~cm$^{-1}$ as a function of gate voltage and the corresponding effective refractive index extracted from the measured wavelength.}
    \label{fig:SI7}
\end{figure}

\newpage
\autoref{fig:SI7} presents additional gate-dependent HP3 measurements in the $5.5^\circ$ twisted bilayer graphene sample. Panel~(a) shows near-field optical amplitude maps acquired at $1480~\mathrm{cm}^{-1}$ for gate voltages ranging from 1 to 10~V, demonstrating gate-tunable HP3 propagation. Panel~(b) shows the HP3 dispersion measured at several gate voltages, revealing the evolution of the polariton wavelength with gate voltage. Panel~(c) summarizes the HP3 wavelength at $1480~\mathrm{cm}^{-1}$ as a function of gate voltage, together with the corresponding effective refractive index.

\vspace{-1em}
\vspace{-1em}

\begin{figure}[H]
    \centering
    \includegraphics[width=0.85\linewidth]{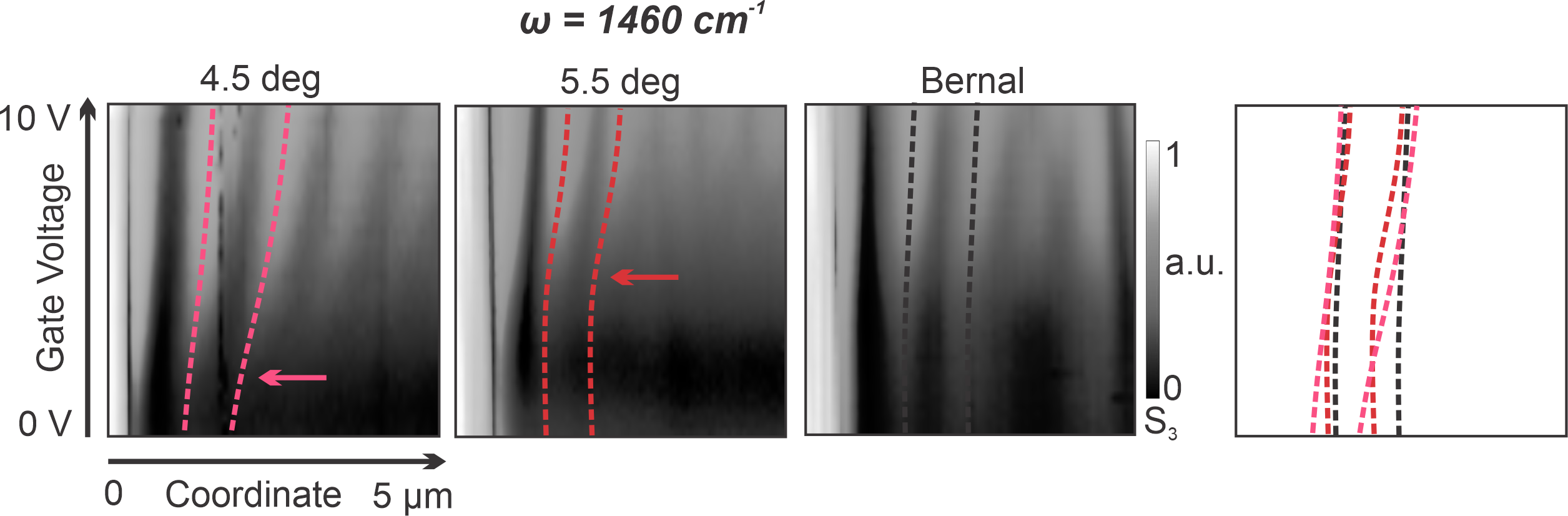}
    \caption{Comparison of HP3 propagation in TBG and Bernal-stacked graphene at $1460~\mathrm{cm}^{-1}$.}
    \label{fig:S8}
\end{figure}

\autoref{fig:S8} compares the near-field optical amplitude associated with real-space HP3 propagation in $4.5^\circ$, $5.5^\circ$, and Bernal-stacked graphene devices at $1460~\mathrm{cm}^{-1}$. In the $4.5^\circ$ and $5.5^\circ$ devices, features associated with the onset of interband transitions are observed as a nonlinear evolution of the HP3 wavelength. In contrast, the Bernal-stacked graphene device exhibits an approximately linear evolution of the HP3 wavelength.

\vspace{-1em}
\vspace{-1em}

\subsection{Temperature dependence of HP3 in $4.5^\circ$ sample at $\omega = 1460~\mathrm{cm}^{-1}$}

\begin{figure}[h!]
    \centering
    \includegraphics[width=0.99\linewidth]{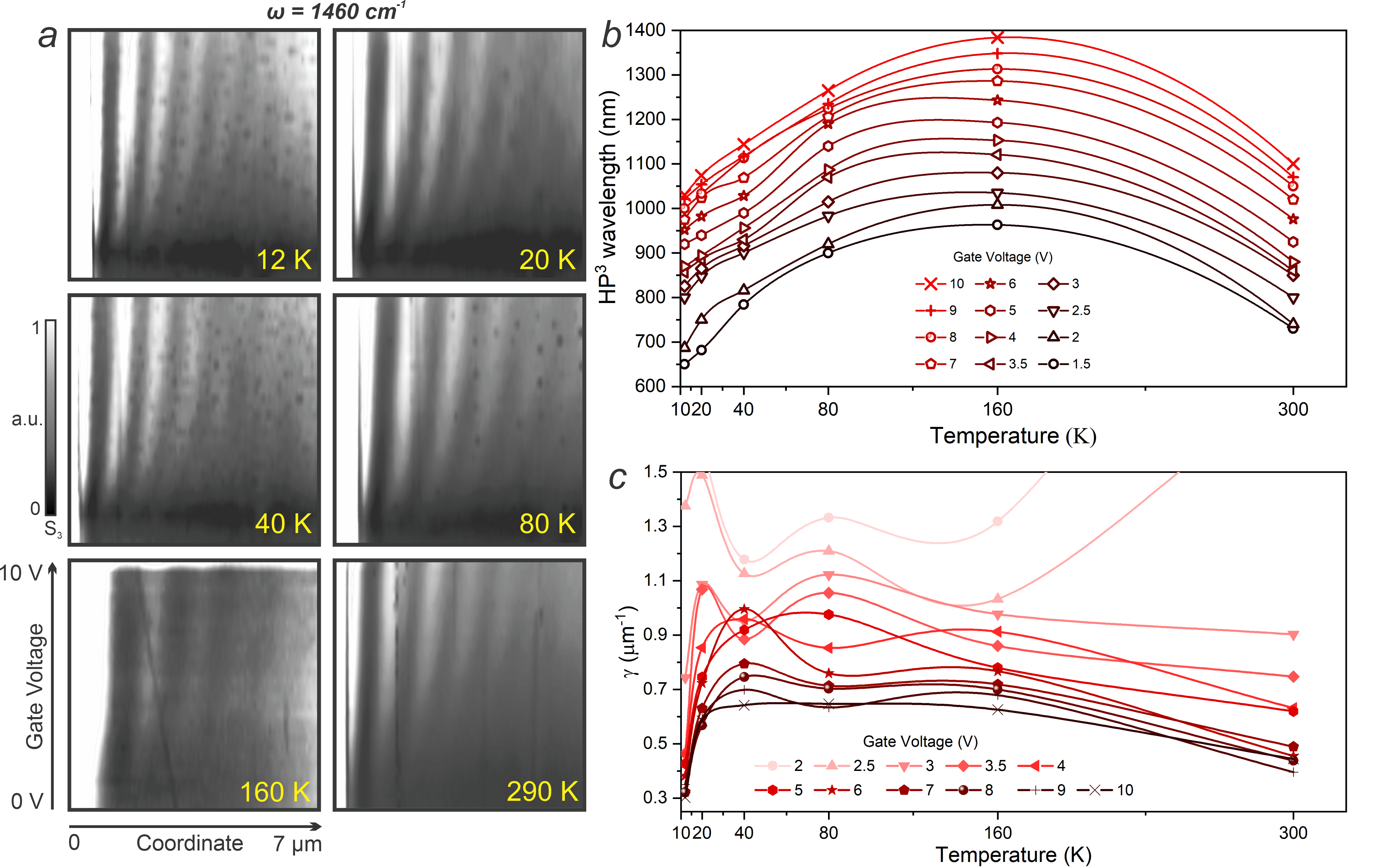}
    \caption{Temperature dependence of HP3 in the $4.5^\circ$ sample at $\omega = 1460~\mathrm{cm}^{-1}$. \textbf{a}) Near-field optical amplitude maps acquired at different temperatures. \textbf{b}) HP3 wavelength as a function of temperature at selected back-gate voltages. \textbf{c}) HP3 damping constant as a function of temperature at selected back-gate voltages.}
    \label{fig:S9}
\end{figure}

\autoref{fig:S9} shows the temperature evolution of the HP3 mode in the $4.5^\circ$ sample at $\omega = 1460~\mathrm{cm}^{-1}$. Panel~(a) shows near-field optical amplitude sweep scans acquired at different temperatures, where the evolution of the HP3 wavelength and damping can be directly observed. Panel~(b) shows the temperature dependence of the HP3 wavelength extracted from line scans at selected back-gate voltages. Panel~(c) shows the temperature dependence of the damping constant extracted from exponential fits to the amplitude decay obtained from the corresponding line scans at selected back-gate voltages.

\newpage
\vspace{-1em}
\section{Hyperbolic plasmon-phonon polaritons in MATBG samples}
\begin{figure}[H]
    \centering
    \includegraphics[width=0.85\linewidth]{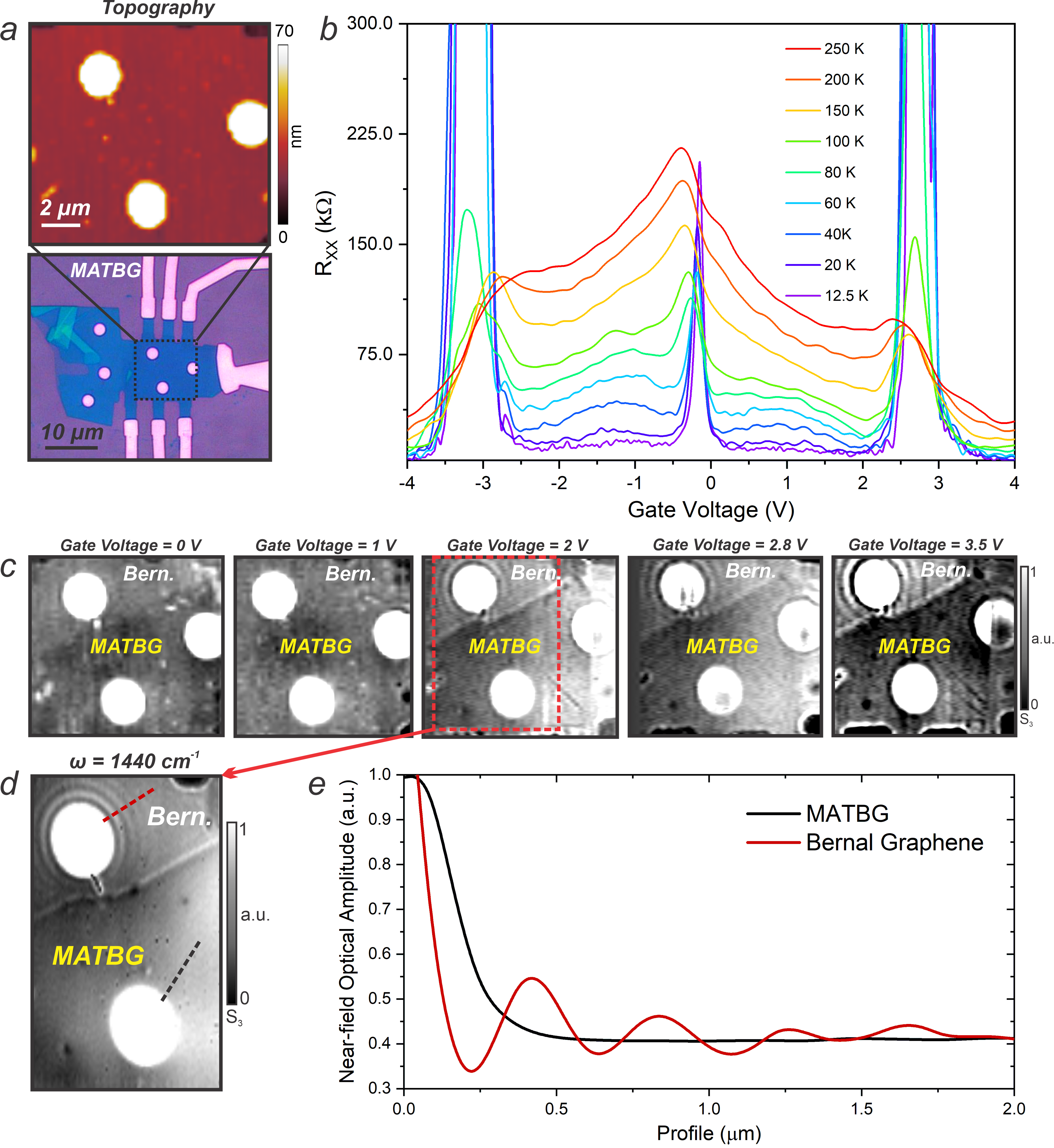}
    \vspace{-1em}
    \caption{Characterization of HP3 propagation in a MATBG device. \textbf{a}) Optical image and AFM topography of the device. \textbf{b}) Longitudinal resistance, $R_{xx}$, as a function of temperature. \textbf{c}) Near-field optical amplitude maps acquired at different back-gate voltages. \textbf{d}) Detailed near-field optical amplitude map acquired at $V_{\mathrm{BG}}=2$~V and $\omega=1440~\mathrm{cm}^{-1}$. \textbf{e}) Near-field optical amplitude profiles extracted from the MATBG (black) and Bernal-stacked (red) regions.}
    \label{fig:S10}
\end{figure}

\vspace{-1em}
An interesting case is the coupling of hyperbolic phonon polaritons to plasmons in a sample close to the magic angle. \autoref{fig:S10}a shows an optical image and AFM topography of the sample. \autoref{fig:S10}b shows the temperature evolution of the longitudinal resistance, $R_{xx}$, from 250~K down to 12.5~K. Band-insulating states can be identified at $\nu=4$. One part of the sample (the top region) consists of Bernal-stacked graphene, whereas the remainder is close to the magic angle.
\vspace{-1em}
\vspace{-1em}
\autoref{fig:S10}c shows near-field optical amplitude maps acquired at different back-gate voltages, corresponding to different regions of the band structure: $V_{\mathrm{BG}}=0$~V is close to the CNP, $V_{\mathrm{BG}}=1$--$2$~V corresponds to the flat-band region, $V_{\mathrm{BG}}=2.8$~V corresponds to the band-insulator peak, and $V_{\mathrm{BG}}=3.5$~V lies beyond the flat-band region in the dispersive band. At gate voltages below approximately 2~V, no HP3 fringes are observed in either the Bernal or MATBG regions, consistent with the low carrier density. At approximately 2~V, HP3 waves emerge in the Bernal-stacked region. At higher gate voltages, HP3 propagation remains visible in the Bernal region but is strongly suppressed in the MATBG region. Notably, this suppression persists even beyond the flat-band region, where the electronic bands become dispersive.

\autoref{fig:S10}d shows a detailed near-field optical amplitude scan acquired at $V_{\mathrm{BG}}=2$~V and $\omega=1440~\mathrm{cm}^{-1}$. \autoref{fig:S10}e shows line profiles extracted from the MATBG (black) and Bernal-stacked (red) regions. The strong suppression of HP3 propagation in the MATBG region over a broad range of carrier densities suggests that the presence of the moir\'e flat bands provides an efficient decay channel for mid-infrared excitations, resulting in enhanced damping of the HP3 mode. This observation warrants further investigation.

\bibliographystyle{naturemag-sergi}
\bibliography{HP3.bib}